\documentclass[journal=jacsat,manuscript=article,layout=onecolumn]{achemso}
\setkeys{acs}{keywords = true}
\usepackage{chemformula} 
\usepackage[T1]{fontenc} 
\usepackage{graphicx}%
\usepackage{multirow}%
\usepackage{amsmath,amssymb,amsfonts}%
\usepackage{amsthm}%
\usepackage{pifont}
\usepackage{mathrsfs}%
\usepackage[title]{appendix}%
\usepackage{xcolor}%
\usepackage{textcomp}%
\usepackage{manyfoot}%
\usepackage{booktabs}%
\usepackage{algorithm}%
\usepackage{algorithmicx}%
\usepackage{algpseudocode}%
\usepackage{listings}%
\usepackage{svg}%
\usepackage{pgfplots}%
\usepgfplotslibrary{statistics}%
\usepackage{tikz} %
\usepackage{listings}%
\usepackage{scontents}%
\usepackage{booktabs}
\usepackage{multirow}%
\usepackage{array} %
\usepackage{nameref}
\usepackage{makecell}
\usepackage{hyperref}
\usepackage{xcolor}
\usepackage[normalem]{ulem}
\usepackage{tabularx}

\pgfplotsset{compat=1.18}
\author{Nuno Costa}
\author{Julija Zavadlav}
\affiliation[TUM]
{Multiscale Modeling of Fluid Materials, Department of Engineering Physics and Computation, TUM School of Engineering and Design, Technical University of Munich, Germany}
\alsoaffiliation[AMC]
{Atomistic Modeling Center, Munich Data Science Institute, Technical University of Munich,
Germany}
\email{julija.zavadlav@tum.de}

\title[Morphology-Aware Peptide Discovery via Masked Conditional Generative Modeling]
  {Morphology-Aware Peptide Discovery via Masked Conditional Generative Modeling}

\abbreviations{CG-MD,AP,SASA,CVAE}

\begin{document}

\begin{tocentry}

\centering
\includegraphics[width=8.25cm,height=4.45cm]{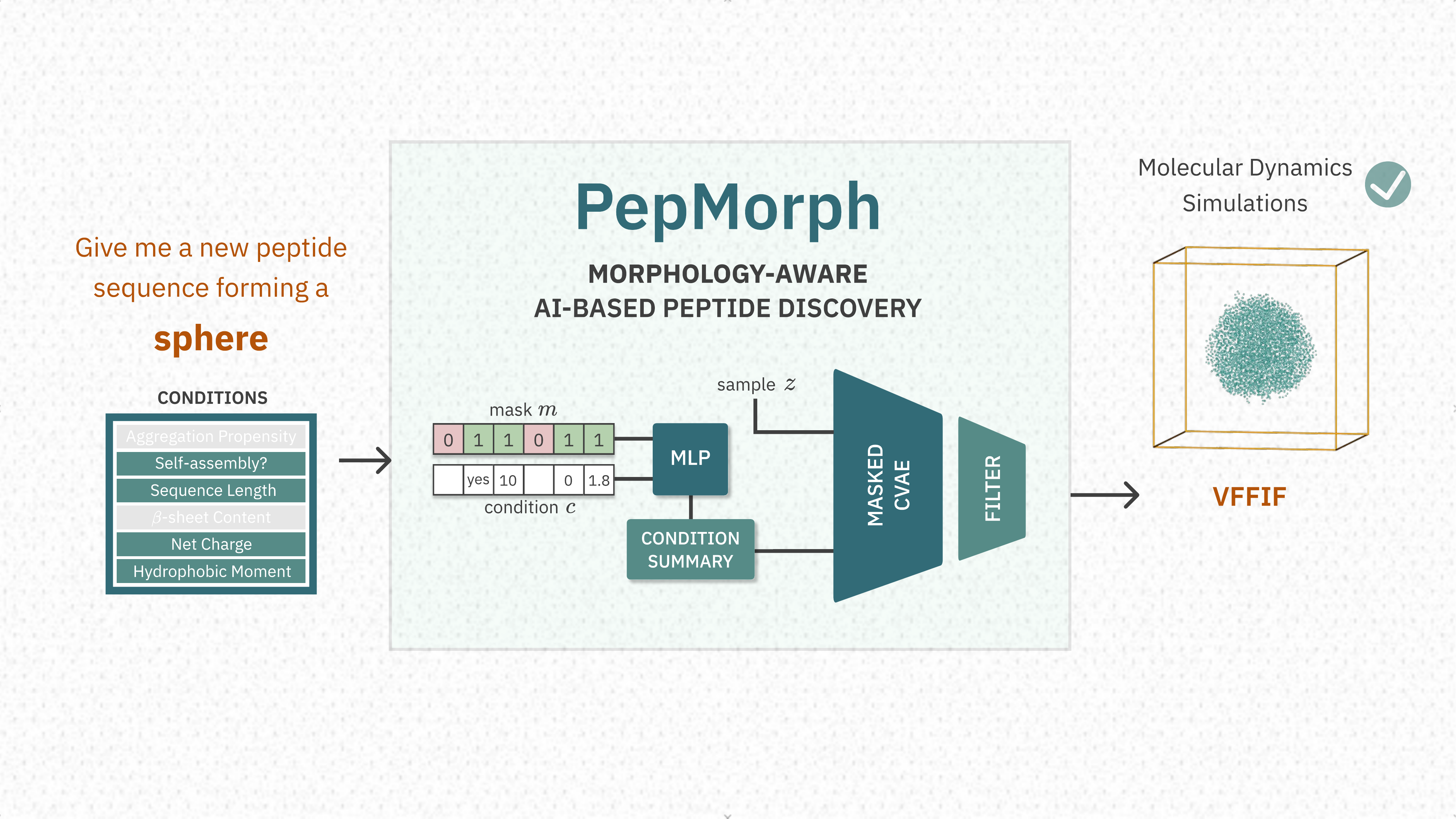}

\end{tocentry}

\begin{abstract}
  Peptide self-assembly prediction offers a powerful bottom-up strategy for designing biocompatible, low-toxicity materials for large-scale synthesis in a broad range of biomedical and energy applications. However, screening the vast sequence space for categorization of aggregate morphology remains intractable. We introduce PepMorph, an end-to-end peptide discovery pipeline that generates novel sequences that are not only prone to aggregate but whose self-assembly is steered toward fibrillar or spherical morphologies by conditioning on isolated peptide descriptors that serve as morphology proxies. To this end, we compiled a new dataset by leveraging existing aggregation propensity datasets and extracting geometric and physicochemical descriptors.  This dataset is then used to train a Transformer-based Conditional Variational Autoencoder with a masking mechanism, which generates novel peptides under arbitrary conditioning. After filtering to ensure design specifications and validation of generated sequences through coarse-grained molecular dynamics (CG-MD) simulations, PepMorph yielded 83\% success rate under our CG-MD validation protocol and morphology criterion for the targeted class, showcasing its promise as a framework for application-driven peptide discovery.
\end{abstract}

\section{Introduction}
\addcontentsline{toc}{section}{\protect\numberline{}Introduction}

Supramolecular self-assembly is a powerful bottom-up strategy for designing functional materials. Small molecular building blocks can spontaneously organize into well-ordered architectures through weak non-covalent interactions (hydrogen bonding, aromatic $\pi$-$\pi$ stacking, hydrophobic effects, electrostatics, metal coordination, etc.). Given the dynamic and reversible nature of these interactions, supramolecular assemblies often exhibit adaptive, self-healing, and stimuli-responsive behaviors~\cite{aidaFunctionalSupramolecularPolymers2012}. These principles have inspired a variety of synthetic supramolecular materials whose structure and function emerge from collective, non-covalent organization.

Among the various supramolecular building blocks, peptides stand out for their inherent biocompatibility, chemical tunability, and straightforward synthesis~\cite{okesolaMulticomponentSelfassemblyTool2018a,sheehanPeptideBasedSupramolecularSystems2021}. The combinatorial diversity afforded by the 20 canonical amino acids defines a vast possible design space for peptide assemblies, as even minor modifications in sequence often yield markedly different supramolecular outcomes, highlighting the critical role of sequence design in defining material structure and function. Indeed, different peptide sequences can form a rich variety of nanostructures --- including fibers, tubes, rods, sheets, vesicles, and micelles --- depending on their primary sequence~\cite{makamMinimalisticPeptideSupramolecular2018a}. At the same time, supramolecular morphology is not purely sequence-intrinsic: assembly outcomes are often co-determined by experimental conditions (e.g., concentration, solvent, pH/ionic strength, temperature) and preparation history, such that the same sequence may yield different morphologies under different contexts~\cite{wehnerSupramolecularPolymerizationKinetic2020, kaygisizContextDependenceAssembly2025a, drazicUsingMachineLearning2025b}.

Peptide-based supramolecular materials have found utility across a diverse array of applications. In biomedicine, for example, peptide assemblies have been employed as drug-delivery vehicles, tissue-engineering scaffolds, biosensors, and theranostic agents ~\cite{guptaUltrashortPeptideSelfAssembly2020a, liPeptidemodulatedSelfassemblyVersatile2019, ashworthPlasticsProteins2022, boddulaPeptidebasedNovelSmall2021, nguyenSelfassemblyDiphenylalaninePeptide2016}. These examples emphasize the potential of peptide self-assembly in life-science and materials-science contexts. However, realizing this potential depends on discovering sequences that adopt target morphologies and satisfy functional requirements, a highly complex challenge given the vast combinatorial sequence space.

Computational design strategies have therefore become necessary for exploring this sequence space. While trial-and-error synthesis is impractical for large-scale screening, conventional rational design, relying on expert intuition and existing heuristics, can be biased and may fail to capture unexpected solutions. In the past few years, machine learning has emerged as a powerful alternative for peptide sequence design, particularly through generative models.
Variational autoencoders (VAEs)~\cite{kingmaAutoEncodingVariationalBayes2022} learn continuous, low-dimensional embeddings of sequences, while conditional VAEs (CVAEs)~\cite{sohnLearningStructuredOutput2015} further enable targeted generation by incorporating property labels or descriptors as conditioning inputs. For instance, PepCVAE demonstrated semi-supervised generation of antimicrobial peptides by conditioning on activity labels \cite{dasPepCVAESemiSupervisedTargeted2018}, as well as the HydrAMP framework built upon this approach to optimize peptide potency and hydrophobic balance for antimicrobial activity \cite{szymczakDiscoveringHighlyPotent2023}. Beyond fixed conditioning, recent CVAE formulations explicitly support arbitrary-subset conditioning by augmenting the conditioning inputs with a binary mask of observed attributes and training the decoder to remain valid when only a subset is provided (e.g., VAEAC)~\cite{ivanovVariationalAutoencoderArbitrary2018}. Related approaches address missing-covariate settings by marginalizing unobserved conditions during training~\cite{ramchandranLearningConditionalVariational2024}.

As it pertains to peptide self-assembly, machine learning has also advanced rapidly, spanning large-scale predictors, autonomous search, and human/experiment-in-the-loop discovery; however, progress remains constrained by data heterogeneity and limited standardized condition-annotated labels~\cite{drazicUsingMachineLearning2025b, batraMachineLearningOvercomes2022, talluriDiscoveryUnconventionalNonintuitive2025, xuAcceleratingPredictionDiscovery2023, yangLearningRulesPeptide2025, wangDeepLearningEmpowers2023a}.
Recently, larger-scale \textit{in silico} screening of the peptide sequence space has enabled the creation of the first extensive aggregation-propensity datasets~\cite{wangDeepLearningEmpowers2023a, liuEfficientPredictionPeptide2023a, vanteijlingenTripeptidesTwoStepActive2021a}. Aggregation Propensity (AP) quantifies the ratio of solvent-accessible surface area (SASA) at the end versus the beginning of a Molecular Dynamics (MD) trajectory~\cite{frederixExploringSequenceSpace2015a}. This metric is a proxy for a peptide's inherent tendency to self-associate and has been used to build classifiers for aggregation given solely the peptide sequences in FASTA format~\cite{liuEfficientPredictionPeptide2023a}. 

Building on these predictors, computational discovery pipelines can propose new sequences and validate them using simulations and/or experiments. Njirjak \textit{et al.} tackle this problem by first training a supervised recurrent neural network classifier on a curated set of 368 peptides and then using it as a fitness oracle within a genetic algorithm to propose sequences with high predicted self-assembly propensity~\cite{njirjakReshapingDiscoverySelfassembling2024a}. Candidates are validated by coarse-grained MD (CG-MD) and, for a subset, experimentally, yielding a reported discovery accuracy of $80-95\%$. However, as with many metaheuristic search approaches, exploration can trade off diversity against exploitation of high-scoring motifs. For example, Njirjak et al. report $>\!40\%$ similarity among generated sequences, which is consistent with convergence toward a small set of high-fitness motifs. This concentration can limit exploration of more distant regions of sequence space and reduce the chance of uncovering alternative self-assembling motifs. Other generative models, as the previously mentioned VAEs and variants, can instead learn broader conditional distributions over sequences --- yielding lower overall similarity while retaining controllability ---, and are therefore promising approaches well suited to discover such alternative motifs.

Crucially, aggregation propensity is only a first filter. Even when aggregation-prone sequences are identified, morphology remains a critical determinant of material function. Applications ranging from drug delivery to biosensors demand specific aggregate geometries (fibers, nets, spheres, etc.), yet datasets linking peptide sequence to computationally or experimentally validated self-assembly morphology are scarce. Recent efforts have started to steer material outcomes such as hydrogelation via human-in-the-loop approaches that couple experiments, simulation, and ML~\cite{xuAcceleratingPredictionDiscovery2023, drazicUsingMachineLearning2025b}; however, generative approaches that can reliably target specific supramolecular morphologies are yet to be explored, especially given that morphology annotations are fragmented across protocols and conditions.
For example, efforts for data gathering concerning morphology have primarily assembled experimental databases such as SAPdb~\cite{mathurSAPdbDatabaseShort2021}, with simulation setups differing substantially across the aggregated studies. This lack of scale and consistency renders the available data unsuitable for training machine learning models. Without direct morphology labels, generative design must rely on indirect proxies as stand-ins to inject structural information; for example, predicted descriptors derived from the peptide sequence and the isolated monomer-level structure model. Features such as monomeric $\beta$-strand propensity, amphiphilicity, and charge distribution were postulated to directly influence whether peptides self-assemble into fibrils, sheets, or globular micelles~\cite{wangAggregationRulesShort2024}. For example, sequences dominated by hydrophobic residues may form compact, spherical aggregates (minimizing exposed hydrophobic surface). Tools like PEP-FOLD \cite{reyPEPFOLD4PHdependentForce2023} allow the effective generation of the most likely 3D conformations of isolated peptides in aqueous solution, which is one of the most common target solvents for supramolecular peptide assemblies.

Computationally derived peptide-level descriptors may, therefore, serve as proxy signals that bias sequence discovery toward assembly behaviors of interest, even if they do not uniquely determine supramolecular morphology. To validate this hypothesis, we introduce PepMorph, a framework designed to condition peptide discovery on aggregation propensity and monomer--level descriptors for morphology awareness. At its core, PepMorph employs a transformer-based CVAE with a masking mechanism, akin to VAEAC, allowing peptide generation to be flexibly conditioned on any set of descriptors of isolated peptides that serve as morphology proxies. We demonstrate that the model generates highly novel and diverse peptides, with $<\!10\%$ similarity, whose properties match all their targets in $55\%$ of cases. All the generated candidates are then passed through a dual-stage filtering pipeline, ensuring that only sequences meeting aggregation and sequence or structure requirements are retained. In end-to-end validation with CG-MD simulations, all sequences formed aggregates, and $83\%$ matched the intended morphology by visual inspection, under our CG-MD validation protocol and morphology criterion, supporting morphology-aware sequence discovery. PepMorph bridges the gap between unguided sequence exploration and fully constrained design, offering a versatile route toward developing functional supramolecular materials.

\section{Results}
\addcontentsline{toc}{section}{\protect\numberline{}Results}

We begin by forming the PepMorph dataset, an extensive aggregation and isolated peptide descriptor dataset. Then, we implement and rigorously evaluate the PepMorph pipeline on penta-to-decapeptide morphology-aware aggregate discovery tasks. To do so, we train a transformer-based CVAE with a masking mechanism on an augmented dataset comprising aggregation metrics and computed descriptors for each peptide sequence. We then demonstrate the model's generative performance by sampling sequences conditioned on assembly and structural metrics. Next, we apply our post-generation filters, selecting fifteen candidates each for fibrillar and spherical morphologies. Finally, we subject these peptides to CG-MD simulations in aqueous solution, quantifying their AP and examining the resulting aggregate morphologies quantitatively and visually.

\subsection*{Dataset Curation and Generation}
\addcontentsline{toc}{subsection}{\protect\numberline{}Dataset Curation and Generation}

\begin{figure}[!ht]
    \centering
    \includegraphics[width=1\linewidth]{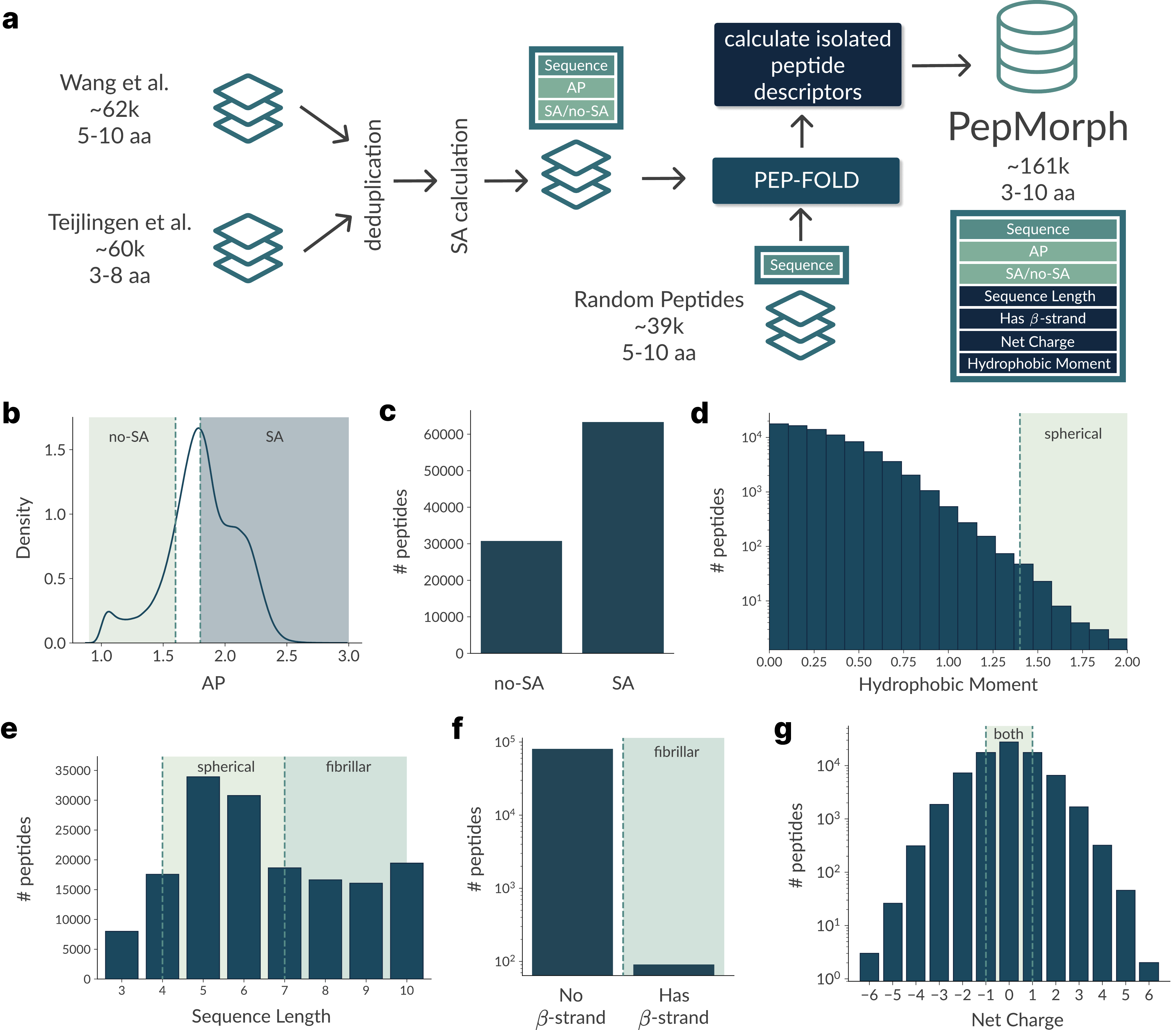}
    \caption{\textbf{PepMorph dataset} (a) Data curation and feature-extraction workflow: we merge three sources: Wang \textit{et al.}~\cite{wangDeepLearningEmpowers2023a} ($\sim$62k peptides, 5-10 amino acids (aa)), Teijlingen \& Tuttle~\cite{vanteijlingenTripeptidesTwoStepActive2021a} ($\sim$60k, 3-8 aa), and a set of $\sim$39k random peptides (retained set from successful PEP-FOLD runs, 5-10 aa). After deduplication, self-assembly (SA) labels are assigned, and peptide conformations are predicted with PEP-FOLD for Wang \textit{et al.} and random peptides to derive biophysical descriptors ($\beta$-strand assignment, net charge and hydrophobic moment). The resulting PepMorph corpus contains ~161k unique peptides spanning 3-10 aa with aggregation-propensity (AP) values and SA/no-SA labels, as well the calculated peptide--level descriptors. Univariate summaries of the PepMorph dataset are shown, specifically of AP density (b), assembly vs no assembly (c), hydrophobic moment (d), peptide length (e), presence of $\beta$-strand assignment (f) and net-charge (g). Regions regarding no-assembly and assembly are highlighted in (b), and condition regions used when targeting specific morphologies are highlighted in the remaining summaries (d-g).}
    \label{fig:dataset}
\end{figure}

One of the main issues with predictive models for short peptides is the lack of extensive datasets like the Protein Data Bank, which provides a comprehensive and standardized structural reference for proteins~\cite{zardeckiPDB101EducationalResources2022}. Nevertheless, we can move towards machine learning models for assembly of short peptides by enriching existing datasets with additional sequences and morphology proxy features (Figure~\ref{fig:dataset}a). 

We curated continuous AP values from Wang \textit{et al.}~\cite{wangDeepLearningEmpowers2023a} and Teijlingen \& Tuttle~\cite{vanteijlingenTripeptidesTwoStepActive2021a}, where AP follows the CG-MD-derived SASA ratio~\cite{wangDeepLearningEmpowers2023a}. Both studies used the MARTINI 2.2 force field. For the 56 peptides common to both datasets, we observed a mean AP difference of $0.089 \pm 0.13$, supporting comparability, and thus mergeability, of the two AP sources (full statistical summary in Supplementary Material). During deduplication, we retained the values reported by Wang \textit{et al.} to maintain internal consistency with the SA/no-SA labeling thresholds and downstream validation protocol adopted throughout this work. Subsequently, we assigned peptide a self-assembly/no-self-assembly (SA/no-SA) flag also based on AP following the operational labeling strategy of Wang \textit{et al.}~\cite{wangDeepLearningEmpowers2023a, liuEfficientPredictionPeptide2023a}: the dataset is split into low-AP vs high-AP classes around the median AP, while excluding samples in a narrow band around the median to reduce ambiguity and label noise. Because the original study reports this procedure but not explicit numeric cutoffs, we reconstructed the corresponding thresholds from their released AP distribution, yielding $\mathrm{AP}\geq 1.8$ (SA) and $\mathrm{AP}\leq 1.65$ (no-SA), with intermediate values left unlabeled (Figure~\ref{fig:dataset}b). These thresholds should, therefore, be interpreted as operational separators rather than actual value--limiters for assembly. Applying this processing step yielded a merged dataset with AP for $121,652$ peptides, $93,668$ with a SA/no-SA label.


Given the masking mechanism of our model, further explained in the next section, PepMorph does not require any specific target to be present to generate sequences (e.g., AP can be absent). As such, we also added random sequences not present in the literature--derived set to expand the conditioning space and give the model broader coverage of the descriptor space. We sampled $50,000$ additional random sequences, by drawing the length uniformly from $3-10$ and sampling each residue independently from the 20 standard aminoacids --- and avoiding duplicates from the Wang \textit{et al.}'s AP data source. This also reduces compositional bias, as the original self-assembly–focused datasets are mildly enriched in aromatic residues, and adding uniformly sampled random sequences shifts the overall amino-acid frequencies closer to uniformity (Supplementary Figure~3). For this random set and for the subset of Wang \textit{et al.}'s peptides that have an SA/no-SA label ($n=41,680$), we predicted isolated 3D conformations with PEP-FOLD 4~\cite{reyPEPFOLD4PHdependentForce2023}, which would then permit the retrieval of the biophysical descriptors that can serve as proxies for aggregate morphology --- computed in the isolated monomeric state and, thus, only used as conditioning signals, not direct aggregation labels. We restricted PEP-FOLD processing of the Wang \textit{et al.}'s set to labeled peptides to focus descriptor extraction on sequences for which downstream assembly-related analyses are defined, and to manage the computational cost of structure prediction. In practice, PEP-FOLD is only applicable to peptides of length $\geq 5$; therefore, synthetic sequences of length $3-4$ were excluded from 3D-descriptor extraction by construction. For the remaining lengths, PEP-FOLD almost always yields a usable lowest-energy model. Rare residual failures did still exist, but are negligible. Even so, this substantially expands coverage of the proxy-descriptor conditioning space (Supplementary Figures~4 and 5). Overall, PEP-FOLD succeeded for $41,402$ ($99.3\%$) SA labeled Wang peptides and for 39,468/50,000 ($78.9\%$) synthetic peptides, yielding 80,870 sequences with valid PEP-FOLD-derived descriptors.

As we will elaborate in the validation pipeline, we focus on three biophysical metrics that plausibly steer spherical versus fibrillar assembly:

\begin{itemize}
\item \textbf{$\beta$-strand assignment}, referring to whether or not any of the amino acid residues adopt an extended strand; peptides with $\beta$-strand propensity tend to stack via backbone hydrogen bonds into elongated, fibrillar (cylindrical) assemblies~\cite{dehsorkhiSelfassemblingAmphiphilicPeptides2014}.
\item \textbf{hydrophobic moment}, measuring the peptide's amphiphilicity; that is, the segregation of hydrophobic and hydrophilic residues in space.
\item \textbf{net charge}, measuring the peptide's overall charge at physiological pH; for example, highly charged peptides experience strong electrostatic repulsion that impedes self-assembly~\cite{dehsorkhiSelfassemblingAmphiphilicPeptides2014}.
\end{itemize}


We obtained valid 3D conformations (and thus retrievable descriptors) for $80,870$ sequences ($\approx 50\%$ of the full corpus). Importantly, this does not mean that the remaining sequences were discarded: PepMorph is trained on the full dataset of $161{,}120$ sequences, described in Table~\ref{tab:descriptor-schema}, with missing descriptor fields (including unavailable 3D descriptors) explicitly represented via the masking mechanism that is explained in the~\nameref{subsec:gen} Section. The distribution across sequence lengths is relatively balanced, with between $15,000$ and $35,000$ peptides represented at each length (Figure~\ref{fig:dataset}e). All possible tripeptides are included, but for longer lengths the dataset covers a smaller fraction of the exponentially expanding theoretical space of $20^L$ sequences (see Supplementary Material). 

\begin{table}[!ht]
\centering
\caption{\textbf{PepMorph dataset composition (161,120 entries)}. It comprises the peptide sequence and length, two aggregation descriptors, and three morphology--proxy descriptors. Coverage counts non-null entries. Synthetic random peptides do not have AP/SA labels by construction; they contribute only sequence-derived fields and the 3D morphology--proxy descriptors. All records are retained for training; unavailable fields appear as missing values and are handled via masking.}
\label{tab:descriptor-schema}
\begin{tabular}{@{}l >{\raggedright\arraybackslash}p{0.25\linewidth} c l@{}}
\toprule
\textbf{Descriptor} & \textbf{Description} & \textbf{Provenance} & \textbf{Coverage} \\
\midrule
\texttt{sequence} & Peptide in FASTA format (uppercase single-letter codes)~\cite{zhanglabFASTAFormat} & \cite{wangDeepLearningEmpowers2023a,vanteijlingenTripeptidesTwoStepActive2021a} + random & \multirow{5}{*}{$161,120$ ($100.0\%$)} \\
\texttt{length}  & Number of residues in \texttt{sequence}                 & derived & \\
\midrule
\texttt{ap} & Aggregation propensity (AP) & \multirow{4}{*}{\cite{wangDeepLearningEmpowers2023a,vanteijlingenTripeptidesTwoStepActive2021a}} & $121,652$ ($75.5\%$) \\
\texttt{is\_assembled} & Self-assembly label (1=SA, 0=no-SA) & & $93,668$ ($58.1\%$) \\
\midrule
\texttt{has\_beta\_strand} & Whether any residue is assigned to $\beta$-strand in the predicted peptide 3D conformation & \multirow{10}{*}{\centering computed} & \multirow{10}{*}{$80,870$ ($50.2\%$)} \\
\texttt{hydrophobic\_moment}   & Magnitude of hydrophobic moment from predicted peptide 3D conformation & & \\
\texttt{net\_charge}           & Peptide net charge at neutral pH & & \\
\bottomrule
\end{tabular}
\end{table}


The resulting PepMorph dataset also exhibits notable imbalances across descriptors. In particular, SA labels are roughly double that of no-SA (Figure~\ref{fig:dataset}c), and $\beta$-strand assignment is exceedingly rare (Figure~\ref{fig:dataset}f). The AP distribution itself is continuous, with many sequences clustering around the cutoff of $\geq\!1.8$ (Figure~\ref{fig:dataset}b). Adding to this, AP shows distinct distributions across lengths (Supplementary Figure~6), showing that the true cutoff can be length--dependent; even so, for simplicity, we keep this threshold across lengths. For these reasons, this threshold should be regarded as an approximate separator. By contrast, continuous descriptors such as the hydrophobic moment (Figure~\ref{fig:dataset}d) and net charge (Figure~\ref{fig:dataset}g) show a broad spread. However, most peptides lie in the low-to-moderate regime for the former. These patterns emphasize the potential bias and structural diversity in the PepMorph dataset. Its associated sparsity is, however, naturally handled by the aforementioned masking mechanism. Furthermore, extending the dataset with additional descriptors from the generated peptide conformations would also be trivial to integrate into the pipeline.

\subsection*{PepMorph Model}
\label{subsec:gen}
\addcontentsline{toc}{subsection}{\protect\numberline{}PepMorph Model}

Incorporating available descriptors into the generative loop introduces a new challenge to a typical Conditional Variational Autoencoder (CVAE). Existing conditional generative models typically require full specification of all conditioning variables during sequence generation. In contrast, a peptide‐design expert may wish to constrain only a subset of properties and leave the rest free. For example, one might enforce a target secondary--structure propensity yet remain agnostic about sequence length or net charge. Rigidly conditioning on every parameter can incorrectly restrict output diversity or prevent generation when specific target attributes are undefined. 

\begin{figure}[!httbp]
    \centering
    \includegraphics[width=\linewidth]{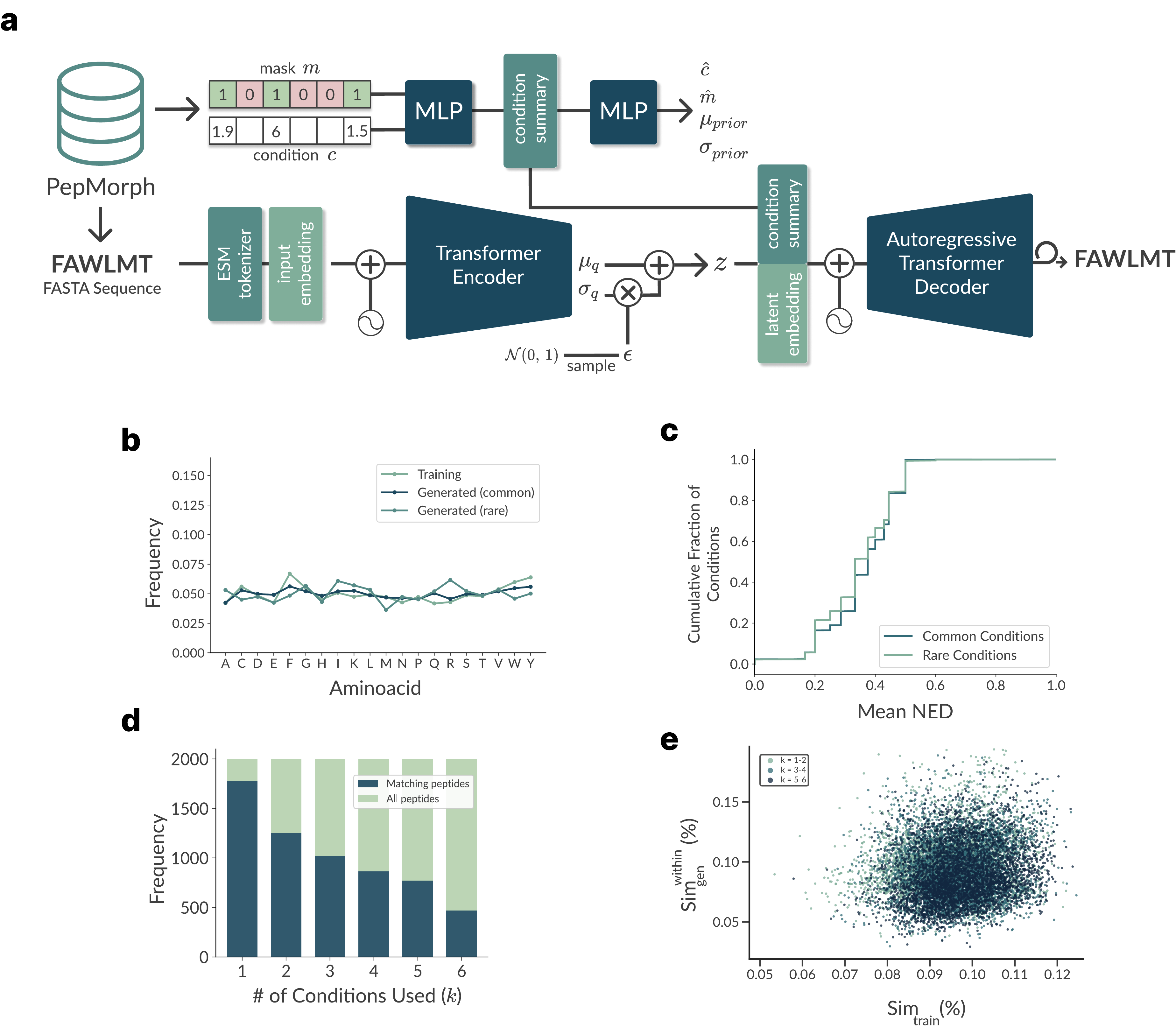}
    \caption{\textbf{PepMorph model and generation validation.} (a) Schematic of the Transformer-based Conditional Variational Autoencoder with the masking mechanism: a descriptor vector $c$ and mask $m$ are summarized into a condition summary that conditions both the latent prior and the autoregressive Transformer decoder, enabling generation under arbitrary subsets of constraints. (b) Amino acid frequency in generated peptides closely follows the training distribution for both common and rare condition sets. (c) Novelty relative to the training set, quantified by the nearest-neighbour normalized edit distance (NED), showing  the empirical cumulative fraction of conditioning targets whose mean NED to the closest training peptide is $\leq x$, which is similar in common vs. rare condition sets. (d) Condition-matching as a function of the number of conditioned descriptors $k$: the fraction of peptides meeting their targets declines as constraints tighten. (e) Similarity via Needleman-Wunsch percent identity of generated sequences (points) to the training set ($\mathrm{Sim}_{\text{train}}$) vs. generated sequences within the same common conditions set ($\mathrm{Sim}_{\text{gen}}^{\text{within}}$), color coded by the number of conditions $k$; values remain near low-identity baselines.}
    \label{fig:model-val}
\end{figure}

To address this limitation and fully leverage our dataset, we develop a transformer-based CVAE that supports \emph{partial conditioning} via an explicit masking mechanism, similarly to the arbitrary--conditioning paradigm of VAEAC --- although in VAEAC this is used for generic feature imputation and image inpainting, rather than targeted masking of descriptors~\cite{ivanovVariationalAutoencoderArbitrary2018} (Figure~\ref{fig:model-val}a). 

Concretely, for $d$ normalized descriptors (min-max scaling using training set statistics) we attach a binary mask $m\in\{0,1\}^{d}$ to the descriptor vector $c\in\mathbb{R}^{d}$ and compute a compact context summary
\begin{equation}
    s \;=\; \phi\!\big([\,c\odot m,\; m\,]\big),
\label{eq:summary}
\end{equation}
where $\odot$ is the element-wise product and $\phi(\cdot)$ is a multilayer perceptron. This summary parameterizes a \emph{masked conditional prior} $p(z\mid s)$ over the latent variable $z$. Intuitively, users fill only the entries they care about; unspecified fields $i$ are masked ($m_i=0$) and are implicitly marginalized by the conditional prior. The decoder then generates the peptide sequence $y$ from $p(y\mid z, s)$. We implement $p$ with an autoregressive Transformer whose cross-attention "memory" contains a latent token (from $z$) and a condition token (from $s$), allowing the model to honor provided constraints while preserving variability (see \nameref{methods}). Our approach mirrors the VAEAC's recipe, conditioning the prior and decoder on the observed context, but uses a learned summary token $s$ instead of injecting raw $[c\odot m, m]$ directly. 

During training, we apply stochastic masking on top of naturally missing descriptors so the model sees arbitrary subsets of observed fields. For each dataset sample, we randomly form a mask $m$, compute $s$ using Eq.~\ref{eq:summary}, and optimize a CVAE objective that encourages the encoder's posterior distribution $q(z \mid y)$ to match the masked conditional prior $p(z \mid s)$ while simultaneously maximizing the likelihood of reconstructing the sequence under the decoder $p(y \mid z,s)$. To ensure that $s$ faithfully encodes the provided context (and only the provided context), we add light auxiliary reconstruction terms: a mask-reconstruction head supervises $m$ with a binary cross-entropy (BCE) loss, two heads supervise the binary descriptors with BCE evaluated only where $m_i=1$, and a small regressor supervises continuous descriptors with a masked mean-squared error (MSE) loss (again only on unmasked dimensions). These terms regularize $\phi$ to capture the observed specifications without imputing missing entries, and empirically improve constraint satisfaction while mitigating posterior collapse. As such, this design accommodates descriptors that are undefined for certain sequences (e.g., PEP-FOLD-derived 3D metrics for very short peptides) and treats them as missing and marginalized rather than imputed. The full loss formulation is shown in the~\nameref{methods} section.

At inference, the user specifies any subset of descriptors (e.g., target length and net charge), sets the corresponding mask entries to $1$, forms $s$, samples $z\sim p(z\mid s)$, and decodes tokens left-to-right until the end-of-sequence marker (\texttt{<eos>}) is yielded.

\subsection*{Conditional Generalization and Sample-based Evaluation}
\addcontentsline{toc}{subsection}{\protect\numberline{}Conditional Generalization and Sample-based Evaluation}

After training, we can discard the encoder component for sampling and leverage the trained mapping of the partial condition sets to the prior distribution and the decoder component as an autoregressive generator. While the generator alone cannot certify assembly or morphology, we can rigorously evaluate (i) \textbf{conditional generalization} on held-out test data, (ii) \textbf{novelty}, measured relative to the training set, (iii) \textbf{diversity} and the \textbf{similarity structure} of the generated set (to verify guidance without collapse), and (iv) \textbf{condition matching} against the requested descriptors. Importantly, these metrics are reported as a descriptive properties of the generated ensembles (e.g., to check against memorization or collapse) and are not inherently used as a performance or quality criterion; functional relevance is defined later.

We firstly report a teacher-forced, token-level negative log-likelihood on a held-out test set as a measure for conditional generalization. Concretely, for each test peptide we condition the decoder on the same subset of observed descriptors available in the dataset, as specified by the mask $m$. As seen in Table~\ref{tab:heldout-ppl}, under this observed-mask conditioning, we obtain a mean per-token cross-entropy of $\approx 0.51$ nats (perplexity $\approx 1.67$). Performance is similar when conditioning on all descriptors for complete-case samples (perplexity $\approx 1.63$) and when applying additional random masking (50\%) on top of the observed fields (perplexity $\approx 1.69$). Overall, these results support that the conditional decoder assigns high likelihood to unseen peptides and that its likelihood remains stable under partial conditioning.

\begin{table}[!ht]
\centering
\caption{\textbf{PepMorph decoder held-out conditional likelihood (teacher-forced).} We report the conditional likelihood on the test set under three masking regimes. In the \emph{observed-mask} case, we evaluate all test samples using all descriptors available for each sample. In the \emph{full-mask} case, we evaluate the subset of the test set for which all descriptors (aggregation and monomer-level) are available. In the \emph{random-on-top} case, we use the same subset as the \emph{full-mask} case but apply additional random masking per sample with probability $p=0.5$ independently for each descriptor. Results are shown as per-token cross-entropy (CE) in nats, with perplexity defined as $\exp(\mathrm{CE})$.}
\label{tab:heldout-ppl}
\begin{tabular}{lcc}
\toprule
Conditioning mask & Per-token CE (nats) & Perplexity \\
\midrule
Observed mask (all test samples) & $\approx 0.51$ & $\approx 1.67$ \\
Full mask (complete-case subset) & $\approx 0.49$ & $\approx 1.63$ \\
Random-on-top (complete-case subset) & $\approx 0.52$ & $\approx 1.69$ \\
\bottomrule
\end{tabular}
\end{table}

Next, we evaluate sample-based generation quality under plausible condition sets. Sampling entirely random descriptor combinations is, however, a poor validation strategy, as arbitrary tuples of values (e.g., conditioning on having $\beta$-strand assignment with sequence length of five) often lie far off the empirical manifold and are either physically inconsistent or statistically implausible, rendering them effectively impossible targets for PepMorph. Instead, we fit a Gaussian Mixture Model per length ($L\in\{5,\dots,10\}$ for comparison with Njirjak \textit{et al.}~\cite{njirjakReshapingDiscoverySelfassembling2024a}) on the training descriptor space and sample condition vectors from these Gaussian Mixture Models, respecting observed correlations and yielding realistic conditions. Because our model supports partial conditioning, we explicitly mask subsets of descriptors at generation time: for each length $L$, we generate 20 conditions, distributing them evenly across $k\in\{1,\dots,6\}$ used descriptors. We also evaluate 10 rare conditions, randomly split across lengths: five with positive $\beta$-strand assignment and five with rare hydrophobic-moment extremes ($>\!0.8$), with all other descriptors set near per-length medians. For each of the $130$ conditions, we decode $100$ peptides autoregressively.

\begin{table}[!ht]
    \centering
    \caption{\textbf{Generation metrics summary.} We report novelty via exact sequence matches to training sequences and the median normalized edit distance (NED) to the most similar (nearest-neighbor) sequence in the training dataset. Diversity is reported via exact sequence matches among newly generated sequences and the pairwise NED across all generated sequences. Similarity is computed using Needleman--Wunsch global alignment with percent identity within training samples, between generated and training sequences, and within generated sequences. Condition matching is reported as the percentage of descriptor matches for the generated sequences. Overall, we observe high novelty, broad diversity, low global similarity, and strong condition fidelity.}
    \label{tab:generation-metrics}
    \begin{tabular}{@{}l r@{}}
        \toprule
        \textbf{Metric} & \textbf{Value} \\
        \midrule
        \multicolumn{2}{c}{\textit{Novelty}}\\
        \midrule
        Exact sequence match & 2.30\% \\
        Nearest-neighbor median NED & 0.3750 \\
        \midrule
        \multicolumn{2}{c}{\textit{Diversity}}\\
        \midrule
        Exact sequence match (mean within conditions) & 0.02\% \\
        Pairwise mean NED (all conditions) & 0.9039 \\
        Pairwise std.\ NED (all conditions) & 0.0928 \\
        \midrule
        \multicolumn{2}{c}{\textit{Similarity (NW \% id.)}}\\
        \midrule
        $\mathrm{Sim}_{\text{train}}$ --- mean   & 9.60\% \\
        $\mathrm{Sim}_{\text{train}}$ --- median & 9.59\% \\
        $\mathrm{Sim}_{\text{gen}}^{\text{all}}$ --- mean   & 9.89\% \\
        $\mathrm{Sim}_{\text{gen}}^{\text{all}}$ --- median & 9.91\% \\
        $\mathrm{Sim}_{\text{gen}}^{\text{within}}$ --- mean   & 9.40\% \\
        $\mathrm{Sim}_{\text{gen}}^{\text{within}}$ --- median & 9.16\% \\
        \midrule
        \multicolumn{2}{c}{\textit{Condition matching}}\\
        \midrule
        Length & 99.72\% \\
        AP     & 81.73\% \\
        SA/no-SA & 84.34\% \\
        Remaining descriptors & 61.12\% \\
        All & 55.21\% \\
        \bottomrule
    \end{tabular}
\end{table}

We quantify novelty in two ways: as exact sequence matching (fraction of generated sequences present in the training set), and as the nearest-neighbor normalized edit distance (NED) (Levenshtein distance divided by the longer length) to the closest training peptide. The PepMorph model generated highly novel sequences (only $\sim 300$ of the 13k generated sequences are present in the training dataset, $97.7\%$ of them are novel) that require, in median, roughly a third of their sequence to be edited to match the closest neighbor on the training set ($0.375$ NED, Figure~\ref{fig:model-val}c). These rare exact matches are expected when sampling from a probabilistic conditional model (and would be far more frequent under memorization). Additionally, the generated peptides' amino acid frequency distribution still closely follows the training set's, and shows no collapse when sampling for rarer conditions (Figure~\ref{fig:model-val}b). 

As for diversity, we assess it with exact sequence matching within the 100 peptides for each generated condition --- showing how well the model explores the space for the same condition set ---, but also from the distribution of pairwise NED \emph{within} each condition (all pairs among the 100 samples) and \emph{all} conditions (100k random pairs of the 13k generated peptides). The results show that the diversity, like the novelty, is also very high (Table~\ref{tab:generation-metrics}): generated peptides are considerably different from each other, evidenced by very high pairwise NED but also extremely low exact sequence matching ($0.02\%$), which indicates that the exploration of the search space can be comprehensive.

To characterize similarity among sequences in a way that is sensitive to conditional structure (and not washed out by an extensive training set), we compute the Needleman--Wunsch global alignment with percent identity~\cite{njirjakReshapingDiscoverySelfassembling2024a} regarding different sets. We report $\text{Sim}_{\text{train}}$, representing the similarity to all training peptides, $\text{Sim}_{\text{gen}}^{\text{all}}$, representing the similarity to all other generated peptides, and $\text{Sim}_{\text{gen}}^{\text{within}}$, representing the similarity to peptides generated from the same condition only. The results (Table~\ref{tab:generation-metrics}) indicate low global percent identity in our sampled ensembles (all similarity means below $10\%$), consistent with broad sampling under partial conditioning. This showcases that a broader conditional distribution was learned, as the method is not bound to a population-based optimization like Njirjak \textit{et al.}'s, which yield substantially higher similarity within their genetic algorithm outputs. From these results, two observations follow: first, $\text{Sim}_{\text{train}}$ staying low argues against memorization; second, $\text{Sim}_{\text{gen}}^{\text{within}}$ remains low when compared to $\text{Sim}_{\text{gen}}^{\text{all}}$, implying that partial conditioning preserves substantial sequence variability within each query rather than collapsing to a few templates. The scatter plot in Figure~\ref{fig:model-val}e also shows a broad cloud of values: we see no significant difference regarding similarity as the number of conditions increases, but we do see some sets of generated peptides with considerably higher similarity (some reaching close to $20\%$ similarity within conditions), indicating that the type of descriptor that it was conditioned on greatly influences the restriction of the manifold. 

Lastly, we quantify condition matching ("effectiveness", Lim \textit{et al.}~\cite{limMolecularGenerativeModel2018}) separately for binary and continuous descriptors. For binary and discrete descriptors, such as SA/no-SA, a sample is a match if the predicted label equals the target. For continuous descriptors, such as AP, a match requires the predicted value to lie within $\pm 10\%$ of the target. We report effectiveness in (Table~\ref{tab:generation-metrics}) both per components --- (i) length only, (ii) AP only, (iii) SA/no-SA only and (iv) all other descriptors --- and as an aggregate score that requires all targeted descriptors to meet their respective criteria simultaneously. This aggregate reflects the model's ability to honor arbitrary subsets of conditioned descriptors. To be able to evaluate the aggregation description matching, we train a separate AP regressor and self-assembly classifier using the PepMorph aggregation-related subset. As detailed in ~\nameref{methods}, we follow Liu \textit{et al.}~\cite{liuEfficientPredictionPeptide2023a}'s approach, but use a single ESM-based transformer backbone~\cite{linLanguageModelsProtein2022a} with the two task-specific heads: an AP regression head and a self-assembly (SA/no-SA) classification head. This unified model yields state-of-the-art performance on both tasks: the AP head attains an MAE of $0.0393$ (matching Liu \textit{et al.}'s $0.0391$), and the SA head reaches $96.72\%$ accuracy (vs. their reported $94.49\%$). 

We predict 3D structures with PEP-FOLD for morphology descriptors and compute the corresponding properties. Overall, condition fidelity is strong ($55\%$ general matching) but descriptor dependent. We see a very high matching rate of length, but considerably lower when measuring against the net charge, $\beta$-strand assignment, or hydrophobic moment. One can also see the monotonic decline of the all-target rate with increasing $k$ used descriptors in Figure~\ref{fig:model-val}d (from $84.58\%$ at $k=1$ down to $24.55\%$ at $k=6$). Even so, high length fidelity alongside solid AP/SA matching indicates that the model is not trading constraint satisfaction for trivial length control; rather, PepMorph can juggle multiple design descriptors without collapsing to a single one. This decline, however, highlights an expected trade-off: as the number of simultaneously fixed descriptors $k$ increases, the probability that a sampled peptide satisfies all requested constraints (under our $\pm10\%$ tolerance for continuous targets) decreases, as any single mismatch causes failure; not only that, but also potentially rarer setting combinations become intrinsically harder to satisfy. In practice, this means PepMorph is most effective when users fix only the few descriptors that are essential to the design intent and leave the remainder unrestrained (via masking), using post-generation screening as a safeguard rather than attempting to "hard-satisfy" many constraints at once.

\subsection*{Morphology Validation via MD Simulations}
\addcontentsline{toc}{subsection}{\protect\numberline{}Morphology Validation via MD Simulations}

To evaluate our framework against the central hypothesis, we deliberately restrict the conditioning descriptors to a small set with values within specific ranges that could act as morphology proxies for two regimes --- spherical and fibrillar, as in Table~\ref{tab:conditions} ---, enriching the probability of obtaining one of the two target morphologies (not uniquely determining them). As such, we perform an end-to-end quantitative evaluation of morphology steering by fixing the condition vector to values in those ranges, but also positively fixing the self assembly descriptor \texttt{is\_assembled} to enforce that aggregation is a prerequisite for morphology formation --- although not a morphology descriptor.  The ranges of the chosen descriptors for each target morphology are duly explained in Table~\ref{tab:conditions}. Because the generator only accepts exact descriptor vectors $c$ but peptide space screening requires ranges, we grid-sweep each descriptor over its specified range with a fixed step size and evaluate all combinations, collecting the corresponding generated peptides. The only aggregation-related input during generation was the binary aggregation label, which we fixed to $1$ for all conditions.

\begin{table}[!ht]
  \centering
  \caption{\textbf{Target property grids for morphology-directed peptide generation.} Hydrophobic moment and net charge are shown in min--max scaled units; targets correspond to high hydrophobic moment and near-neutral net charge. Combinatorial condition grids were sampled with increments of $1$ (length), $0.1$ (hydrophobic moment), and $0.05$ (net charge). We treat these ranges as heuristic steering windows rather than deterministic rules, and validate morphology only via downstream CG-MD later on.}
  \label{tab:conditions}

  \begin{tabular}{@{} 
      >{\raggedright\arraybackslash}p{2cm}
      l
      c 
      >{\raggedright\arraybackslash}p{7.5cm}
    @{}}
    \toprule
    \textbf{Target Assembly} & \textbf{Property} & \textbf{Range} & \textbf{Rationale} \\
    \midrule
    \multirow{3}{=}{Spherical Aggregates}
      & Peptide Length       
      & $5-7$aa
      & Short peptides are less likely to adopt extended conformations that are characteristic of elongated aggregates~\cite{stangerLengthdependentStabilityStrand2001}. \\
      & Hydrophobic Moment
      & $0.6-1.0$
      & Moderate amphipathicity favors the burial of hydrophobic faces 
        leading to amphiphiles like vesicles or micelles
        \cite{familiaPredictionPeptideProtein2015}. \\
      & Net Charge            
      & $0.4-0.6$
      & Near-neutral charge reduces repulsion, enabling self-assembly  
        \cite{dehsorkhiSelfassemblingAmphiphilicPeptides2014}. \\
        
    \midrule
    \multirow{4}{=}{Fibrillar Aggregates}
      & Peptide Length        
      & $7-10$aa
      & Longer peptides may adopt extended conformations that can align into $\beta$-sheets 
        \cite{stangerLengthdependentStabilityStrand2001}. \\
      & Has $\beta$-strand
      & Yes
      & High monomeric $\beta$-strand assignment drives backbone H-bond stacking into longer fibrils 
        \cite{dehsorkhiSelfassemblingAmphiphilicPeptides2014}. \\
      & Net Charge              
      & $0.4-0.6$
      & Near-neutral charge reduces repulsion, enabling self-assembly
        \cite{dehsorkhiSelfassemblingAmphiphilicPeptides2014}. \\
    \bottomrule
  \end{tabular}
\end{table}

\begin{figure}[!ht]
    \centering
    \includegraphics[width=\linewidth]{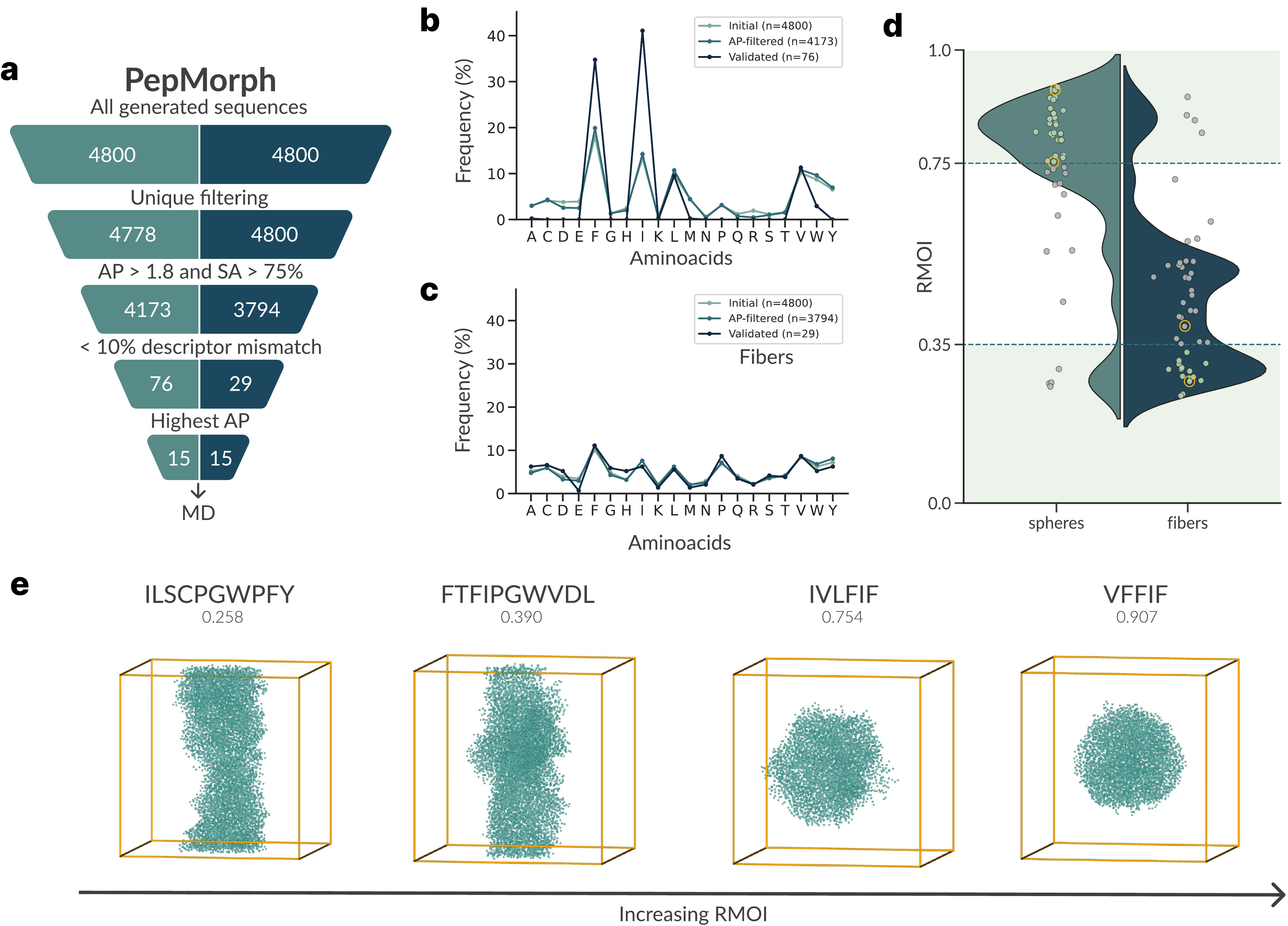}
    \caption{\textbf{PepMorph pipeline for spherical vs. fibrillar aggregate generation: screening and Molecular Dynamics (MD) visualization.} (a) Screening funnel for the two targeted morphologies (left values refer to spheres, right values to fibers).
    Amino acid occurrence across the funnel for (b) spheres and (c) fibers. For spheres, the validated set collapses to a narrow alphabet dominated by F/I/L/V, whereas fibers remain compositionally closer to the pre-filter pool. (d) Distributions of the RMOI for all MD runs (3 per selected peptide, leading to 90 runs); dashed lines mark success thresholds (spheres $\geq\!0.75$, fibers $\leq\!0.35$), with individual peptide points overlaid. Representative MD snapshots are circled in (d) and shown in (e), with the corresponding sequence and RMOI above each panel illustrating the progression from fibrillar to spherical aggregates.}
    \label{fig:sim-val}
\end{figure}

Despite the flexibility afforded by partial conditioning, the stochasticity of generative sampling and the additional noise introduced by masking imply that the generated sequences may still deviate from their target properties. To enforce fidelity to the design criteria, we incorporate a post-generation filtering stage using established predictors (Figure~\ref{fig:sim-val}a). 

First, sequences are screened with the aforementioned AP regressor and SA/no-SA classifier to remove sequences unlikely to aggregate. Next, each candidate peptide is fed into PEP-FOLD to obtain its 3D conformation; from it, together with the sequence, we can recompute all peptide-level descriptors. Only those whose predicted metrics fall within the predefined tolerance ($\pm 10\%$) of the conditioned values are retained. This dual filtering step transforms our pipeline into a reliable end-to-end framework that yields peptides meeting both aggregation and proxy specifications.

To keep a matched sampling budget across the two target regimes, we generated a fixed total of $4,800$ sequences per target. Given that the condition grids have a different number of possible combinations, this corresponded to $80$ distinct peptides per condition for spherical targets and $300$ for fibrillar targets. After deduplication, the spherical set decreased to $4,778$ sequences, whereas the fibrillar set remained at $4,800$ (no duplicates). We then applied our post-processing filter to prioritize highly aggregating candidates: we retained only peptides with predicted aggregation probability $\ge\!75\%$ from the SA/no-SA classifier, and with predicted aggregation propensity (AP) $\geq\!1.8$ from the regressor. Following this initial screening, $4,173$ spherical and $3,794$ fibrillar sequences remained. After the aggregation-based screening, we assess peptide-level descriptor compliance and predict structures for all remaining sequences with PEP-FOLD, discarding any peptide that fails to satisfy all targeted/conditioned descriptors within the $10\%$ threshold. This yields $76$ spherical candidates and $29$ fibrillar candidates. From the combined candidate set (both morphologies), we select the top $15$ sequences for each morphology by predicted aggregation propensity (AP) from our trained AP regressor, and proceed with validation via CG-MD simulations.

To verify whether sequences that match the proxies arise from conditioning during generation rather than post-generation rejection sampling alone, we also evaluated two baselines under the same screening pipeline and the same sampling budget (4,800 sequences per target regime). First, we generated 4,800 uniformly random peptides and applied the full aggregation and descriptor filtering pipeline; this yielded \textbf{no} spherical candidates and only \textbf{4} fibrillar candidates. Second, we used PepMorph in a \emph{length-only} mode by masking all morphology-proxy descriptors at generation time and relying solely on post-generation screening; under the same budget, this yielded only \textbf{3} spherical candidates and \textbf{7} fibrillar candidates. These results, explicitly shown in Table~\ref{tab:baseline_yields}, indicate that sampling without morphology conditioning is markedly sample-inefficient, and that conditioning during generation is essential to enrich the candidate pool before MD validation. Because PepMorph is a sampling-based constrained discovery pipeline rather than an iterative optimization loop, we quantify efficiency by the probability of obtaining a proxy-compliant candidate (or a "proxy-hit") under a finite screening budget, which directly implies the expected number of screening calls per hit. We report corresponding Wilson 95\% confidence intervals, calls-per-hit estimates and budget--success curves, as well as the full funnels for both of these baselines, in Supplementary Material.

\begin{table}[!ht]
\centering
\caption{\textbf{Baselines for morphology-targeted discovery via post-filtering.}
Under the same sampling budget ($N=4{,}800$ generated sequences per target regime) and the same post-generation screening pipeline, conditioning on morphology-proxy descriptors substantially increases the yield of proxy-compliant candidates. Calls-per-hit are computed as $N/\#\text{hits}$ (infinite if zero hits).}
\label{tab:baseline_yields}
\small
\setlength{\tabcolsep}{8pt}
\renewcommand{\arraystretch}{1.25}
\begin{tabular}{lcc}
\toprule
\textbf{Generator / sampling method} &
\textbf{Sphere-target} &
\textbf{Fibril-target} \\
\midrule

PepMorph model, all proxies &
\makecell[c]{76 / 4{,}800\\\footnotesize($\approx 63$ calls/hit)} &
\makecell[c]{29 / 4{,}800\\\footnotesize($\approx 166$ calls/hit)} \\

PepMorph model, length-only &
\makecell[c]{3 / 4{,}800\\\footnotesize($1600$ calls/hit)} &
\makecell[c]{7 / 4{,}800\\\footnotesize($\approx 686$ calls/hit)} \\

Random sequences (uniform residues) &
\makecell[c]{0 / 4{,}800\\\footnotesize($\infty$ calls/hit)} &
\makecell[c]{4 / 4{,}800\\\footnotesize($1200$ calls/hit)} \\

\bottomrule
\end{tabular}
\end{table}

For the chosen set of peptides for CG-MD validation, we can quantitatively distinguish both morphologies from the resulting MD trajectory via the ratio of principal moments of inertia (RMOI) introduced by Wang \textit{et al.}~\cite{wangAggregationRulesShort2024}. For the largest aggregate cluster, let $\lambda_1 \leq \lambda_2 \leq \lambda_3$ denote the eigenvalues of its inertia tensor (see Methods); then
\begin{equation}
    \mathrm{RMOI} = \frac{\lambda_1}{\lambda_3}.
    \label{eq:RMOI}
\end{equation}
By construction, $\mathrm{RMOI} \in (0,1]$: elongated, fibrillar aggregates yield values closer to $0$, whereas compact, spherical aggregates approach $1$. Following Wang \textit{et al.}, $\mathrm{RMOI} \leq 0.35$ can be classified as fibrillar or tubular, while $\mathrm{RMOI} \geq 0.75$ denotes spherical assemblies. Intermediate values fall into an undefined regime that may correspond to amorphous aggregates or diverse morphologies such as sheets or nets.

All 30 peptides yielded by PepMorph for simulation-based validation, as well as additional control cohorts totalizing 87 peptides, were simulated in triplicate under the same CG-MD protocol and analysis pipeline. For PepMorph-selected peptides, aggregation was observed in every run: each exceeded the aggregation-propensity threshold ($\mathrm{AP} \geq 1.8$, computed from the CG-MD SASA ratio) and showed visible assembly upon trajectory inspection. With respect to morphology targeting, the RMOI distributions show distinct tendencies for the two classes: spherical assemblies cluster around high values ($\sim 0.8$), while fibrillar assemblies concentrate at low values ($\sim 0.3$), as seen in Figure~\ref{fig:sim-val}d,e. This separation highlights that RMOI captures the expected contrast between compact and elongated morphologies, albeit only as a proxy. Nevertheless, the cutoff values were empirically defined, and the metric has significant limitations. We observed several borderline cases where RMOI classified assemblies incorrectly, despite visual inspection confirming the target morphology. Because RMOI considers only the longest and shortest inertia directions, it neglects how mass is distributed in the simulation box --- for instance, aggregates aligned along box boundaries in all directions sometimes yielded high RMOI despite being fiber-like. More broadly, RMOI struggles to reliably classify fibrils: slightly wider fibers tend to exhibit intermediate values, and the metric inherently overemphasizes tubular/rod geometries rather than fibrillar ones. By visually inspecting the final frame of each trajectory (see Supplementary Material), we determine a morphology success rate under our CG-MD validation protocol of 83\% (Table~\ref{tab:success}). Visual labeling used explicit geometric criteria: spherical denotes a single compact aggregate with low apparent anisotropy, whereas fibrillar denotes an elongated aggregate with a dominant long axis; ambiguous or sheet-/net-like assemblies were labeled as failures for the targeted class.

As previously mentioned, and to contextualize this CG-MD-based validation and provide negative baselines, we additionally selected multiple control cohorts. These isolate (i) aggregation enrichment from the AP/SA screening stage and (ii) morphology steering from the proxy-descriptor windows: we selected random peptides (n=15) and non-morphology-proxy (unconditional) generated peptides (n=12; two per length), which provide lower-bound baselines with no design intent; we then selected cohorts that failed the AP filter (AP-fail), which test whether aggregation is obtained without the aggregation screen (n=30; n=15 per morphology); and then cohorts that passed the AP filter but failed the descriptor match, which test whether aggregation alone is sufficient to recover the intended morphologies without morphology-proxy constraints (n=30; n=15 per morphology).

As summarized in Table~\ref{tab:success}, AP-fail cohorts aggregate rarely, while descriptor-fail cohorts often aggregate but achieve the intended morphologies at markedly lower rates than PepMorph candidates, indicating that aggregation enrichment and morphology steering are separable effects in the pipeline. To quantify the strength and uncertainty of morphology steering, we compare PepMorph candidates against the aggregation-matched descriptor-fail controls using Wilson 95\% confidence intervals for success rates and two-sided Fisher exact tests, reported in Supplementary Material. By visual criteria, PepMorph exhibits substantially higher morphology success, with odds ratios of $2.5$ ($p=0.410$) for spheres and $7.6$ ($p=0.042$) for fibrils. If we pool across the two targeted tasks, it yields an odds ratio of $4.3$ ($p=0.022$). These results support that the proxy-descriptor windows enrich for the intended morphologies beyond aggregation screening alone, while the modest MD cohort sizes imply wide uncertainty, particularly for the spherical target.

\begin{table}[!ht]
  \centering
    \caption{\textbf{CG-MD aggregation and morphology outcomes for PepMorph candidates and control cohorts.} Aggregation is defined by AP $\geq 1.8$ (majority across three runs). For targeted cohorts, morphology success rates (RMOI and visual) are reported conditional on aggregation. RMOI success uses a majority vote across three runs given thresholds (spheres $\geq 0.75$, fibrils $\leq 0.35$), and visual success is also based on majority vote across three runs. For untargeted cohorts (random/unconditional), no morphology target is defined; we therefore report, among \emph{aggregating} sequences, the fraction whose aggregates fall into sphere-like (S) or fibril/tube-like (F) regimes.}
  \label{tab:success}

  \small
  \setlength{\tabcolsep}{3.5pt}
  \renewcommand{\arraystretch}{1.08}

  \begin{tabularx}{\linewidth}{@{}p{6cm}c>{\centering\arraybackslash}X>{\centering\arraybackslash}X@{}}
    \toprule
    \textbf{Cohort} &
    \makecell[c]{\textbf{Aggregation}\\\textbf{rate}} &
    \makecell[c]{\textbf{RMOI}\\\textbf{outcome}} &
    \makecell[c]{\textbf{Visual}\\\textbf{outcome}} \\
    \midrule

    \multicolumn{4}{@{}l}{\textbf{Untargeted}} \\
    Random (n=15) &
      20.0\% (3/15) &
      \makecell[c]{S: 0.0\% (0/3)\\F: 0.0\% (0/3)} &
      \makecell[c]{S: 0.0\% (0/3)\\F: 0.0\% (0/3)} \\
    Unconditional (n=12) &
      58.3\% (7/12) &
      \makecell[c]{S: 14.3\% (1/7)\\F: 57.1\% (4/7)} &
      \makecell[c]{S: 14.3\% (1/7)\\F: 71.4\% (5/7)} \\
    \midrule

    \multicolumn{4}{@{}l}{\textbf{Spherical-targeted}} \\
    Failed AP screen (n=15) &
      6.7\% (1/15) &
      0.0\% (0/1) &
      0.0\% (0/1) \\
    Failed descriptor screen (n=15) &
      86.7\% (13/15) &
      46.2\% (6/13) &
      61.5\% (8/13) \\
    PepMorph candidates (n=15) &
      \textbf{100.0\% (15/15)} &
      \textbf{73.3\% (11/15)} &
      \textbf{80.0\% (12/15)} \\
    \midrule

    \multicolumn{4}{@{}l}{\textbf{Fibril-targeted}} \\
    Failed AP screen (n=15) &
      26.7\% (4/15) &
      0.0\% (0/4) &
      50.0\% (2/4) \\
    Failed descriptor screen (n=15) &
      86.7\% (13/15) &
      23.1\% (3/13) &
      46.2\% (6/13) \\
    PepMorph candidates (n=15) &
      \textbf{100.0\% (15/15)} &
      \textbf{26.7\% (4/15)} &
      \textbf{86.7\% (13/15)} \\

    \midrule
    \textbf{PepMorph (all)} (n=30) &
      \textbf{100.0\% (30/30)} &
      \textbf{50.0\% (15/30)} &
      \textbf{83.3\% (25/30)} \\
    \bottomrule
  \end{tabularx}
\end{table}

We also performed targeted robustness controls, fully displayed and contextualized in Supplementary Material, on a randomly selected subset of validated peptides (8 per target; triplicate) to probe sensitivity to to force-field choice in aggregation propensity (MARTINI 2.2 versus MARTINI 3), and periodic-boundary artifacts for morphology (larger box at matched concentration). Aggregation propensity remains in the same regime across setups, supporting that aggregation propensity is comparatively transferable across MARTINI 2.2 and MARTINI 3. In contrast, morphology is more sensitive: spherical outcomes remain largely consistent under the larger-box condition, whereas fibrillar outcomes show some setup sensitivity, yielding networked or amorphous aggregates. Some of these results could still, however, be attributed to short and yet-to-converge trajectories, especially for the larger box, which emphasizes the extensive computational cost associated with larger setups. Even so, this motivates interpreting the MD-based morphology validation as protocol-conditional and highlights that proxy selection and validation protocol jointly determine end-to-end outcomes.


As it pertains to the overall throughput of the pipeline, compared with validation under training-like conditions using our Gaussian Mixture Model procedure, enforcing the target morphology descriptors leads to a substantially larger drop in valid generated peptides. This suggests that the chosen condition combinations are both underrepresented in the training distribution and intrinsically difficult, yielding low data support and fewer sequences that satisfy all constraints. The effect is particularly pronounced for fibrillar targets: the required $\beta$-strand assignmet occurs in $<\!0.1\%$ of training peptides (Figure~\ref{fig:dataset}f), which further depresses the hit rate.

\begin{figure}[!ht]
    \centering
    \includegraphics[width=1\linewidth]{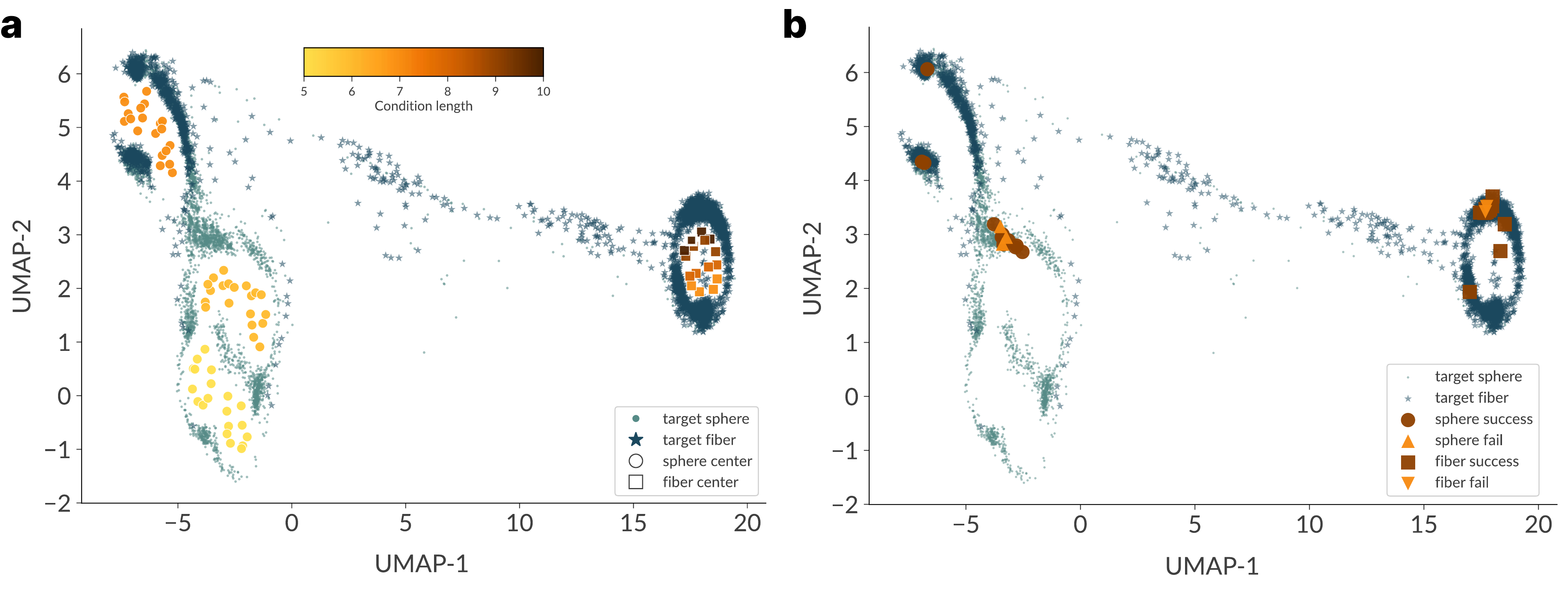}
    \caption{\textbf{Latent map of morphology conditioning.} A UMAP is fit on the conditional prior centers with encoder posteriors of generated peptides projected into the same space (dots for spheres, stars for fibers). (a) Condition center embeddings are colored by conditioned sequence length, revealing an ordered length gradient along the sphere branch and a single, compact, and distinct fiber island; notably, fiber targets of length 7 cluster closer to the sphere branch, consistent with over-constrained fiber conditions claims. (b) Same embedding with MD-simulated peptides highlighted according to target morphology and outcome (success/failure under the RMOI criterion); stars mark the sphere/fiber prior centers.}
    \label{fig:results}
\end{figure}

We also fit a Uniform Manifold Approximation and Projection (UMAP) on the conditional prior centers and project all generated peptide embeddings (from the encoder) into this latent space (Figure~\ref{fig:results}). 

This visualization contextualizes the morphology-specific conditioning: fiber-targeted conditions collapse into a compact, sparsely populated island, consistent with the rarity of $\beta$-strand-positive peptides, whereas sphere-targeted conditions span several sub-modes, reflecting broader but structured support. Even so, it is important to recall that the validated sphere set remains compositionally narrow, with amino acid usage dominated by hydrophobic residues like F/I/L/V (Figure~\ref{fig:sim-val}b), suggesting that the descriptor choices together with the AP $\geq\!1.8$  and structural filters bias toward a restricted set of chemistries. In particular, an atomistic follow-up on two highly hydrophobic sphere-target sequences, present in Supplementary Material Section \textit{Targeted fully atomistic re-simulations}, results in fibrillar aggregation, while all CG-MD runs --- including with the MARTINI~2.2 and larger box runs --- yielded a spherical aggregate. This illustrates that high hydrophobic moment alone does not prevent fibrillar outcomes when overall hydrophobicity is extreme, motivating future proxy refinement. A closer look at the fiber samples reveals a connection between the sequence length of seven in the region corresponding to spheres and the island corresponding to true fiber conditions. This suggests that conditioning for shorter sequences with $\beta$-strand assignment, which was used as a target for fibers, is likely an overconstrained requirement (possibly unrealistic), as the resulting sequences do not cluster near their intended condition centers. This reinforces both the challenges of overconstraining and the importance of masking and range exploration in maintaining valid generative support.








\section{Discussion}
\addcontentsline{toc}{section}{\protect\numberline{}Discussion}

We have developed PepMorph, an end-to-end framework for generating novel, diverse, aggregation-prone peptides, explicitly conditioned on geometric and physicochemical descriptors that inform their assembly morphology. Our pipeline is extremely successful at generating highly novel and diverse sequences, with $83\%$ of them adopting the intended morphology under targeted conditioning. PepMorph addresses a gap in peptide generative modeling by enabling user-specified proxy-guided morphology control through a transformer-based CVAE with a masking mechanism for flexible conditioning on peptide descriptors, whereas prior peptide generative models have predominantly focused on functional activities such as antimicrobial properties~\cite{dasPepCVAESemiSupervisedTargeted2018, szymczakDiscoveringHighlyPotent2023}, and existing self-assembly-oriented generators have not provided any morphology steering as part of the design objective~\cite{njirjakReshapingDiscoverySelfassembling2024a}.

A key advantage of PepMorph is that its conditioning mechanism is both descriptor-agnostic and mask-aware: any peptide descriptor with an assignable value (and, optionally, a predictor for validation) can be integrated without changing the core architecture. This makes the framework naturally extensible along two axes. First, it can easily scale in the sequence space: extending to longer peptides primarily requires additional data and model capacity, while the masked conditioning and autoregressive decoding remain directly applicable. Second, it can expand in the morphology space: the same interface can accommodate alternative morphologies by using different and possibly richer or higher-fidelity descriptors, enabling targets beyond spheres and fibrils (e.g., sheets or nets). The main challenge here lies in quantitatively characterizing more complex morphologies --- an inherently difficult task, as illustrated by the limitations of RMOI. Beyond morphology, application-driven properties such as antimicrobial activity, previously used in generative peptide design~\cite{liFoundationModelIdentifies2024, szymczakDiscoveringHighlyPotent2023}, can also be seamlessly incorporated. In short, PepMorph provides a reusable scaffold for expanding from short peptides and two morphologies to broader sequence lengths and richer structural or functional targets.

While the current descriptor set demonstrates morphology steering under our chosen targeting window, two practical issues are worth improving. First, the choice of proxy descriptors, as well as their feasible target ranges, materially affect end-to-end success. In the present work, morphology control is therefore best interpreted as protocol-conditional: it is validated under a fixed CG-MD setup and morphology criterion, and additional sensitivity checks indicate that outcomes can shift under protocol changes (e.g., larger boxes) or higher-resolution atomistic settings for specific sequences. This highlights that expert-guided refinement of the proxy set --- for example, adding complementary constraints such as an explicit hydrophobicity cap for sphere targeting --- may improve robustness without altering the core partial-conditioning framework. Second, some regions of descriptor space are considerably underrepresented (e.g., sequences with any $\beta$-strand assignment are $\sim0.1\%$, with almost all of them having 10 residues --- see Supplementary Material), which can bias learning and make those targets harder to satisfy. Curating additional data in these sparse regimes and performing expert-guided descriptor selection are straightforward, high-yield next steps that complement the existing pipeline.

Regarding the model, the simple approach of modeling the conditional prior as a single diagonal Gaussian works well in our setting, as reflected by strong condition matching performance in validation. However, under partial conditioning with a sparse mask $m$, the mapping from descriptors to valid sequences is inherently multimodal. A unimodal prior averages across disparate modes, which weakens constraint satisfaction and can cause the decoder to default to frequent training patterns. A promising direction for future work would be to develop architectures that adopt a richer masked prior, such as explicitly multimodal, mixture-based or flow-based prior formulations~\cite{wangDiverseAccurateImage2017, lavdaImprovingVAEGenerations2019, leeDesignPeptidesNoncanonical2025}, making the uncertainty induced by masked descriptors explicit, rather than collapsing it into a single Gaussian.

Since the aggregation labels in our dataset were obtained with CG-MD, we validated candidate morphologies \textit{in silico}, using a similar CG-MD setup, rather than experiments. This choice ensures methodological consistency, but MD predictions are known to be sensitive to the employed force field and simulation setting, and do not always match the experimental predictions. Thus, re-labeling the aggregation labels with more accurate molecular models would be very impactful. Particularly promising in this respect are machine-learning interatomic potentials~\cite{thalerLearningNeuralNetwork2021,rockenAccurateMachineLearning2024} --- especially implicit-solvent variants~\cite{rockenPredictingSolvationFree2025,costeDevelopingImplicitSolvation2024,thalerDeepCoarsegrainedPotentials2022} --- which can approach ab-initio accuracy while retaining near-linear scaling with system size. Integrating such models would yield higher-fidelity training and validation labels while keeping the end-to-end design cycle computationally tractable. In addition, the simulation-generation loop could be tightened by calculating and conditioning directly on morphology descriptors such as RMOI. Ultimately, more extensive atomistic simulations or even experimental characterization would be required to establish morphology as ground truth and to calibrate which proxy descriptors and simulation protocols transfer reliably across conditions. This would be a natural next step to extend and further validate this work.

In summary, PepMorph represents a versatile generative framework for peptide self-assembly that enables treating morphology as an explicit design objective through partial conditioning. Coupled with an MD-validated screening loop, it yields low-redundancy, morphology-specific candidates with measurable shape outcomes, indicating its capability of navigating the vast peptide sequence space. We see this work as a stepping stone toward truly designable peptide assemblies, bridging the gap between sequence specification and material form.

\section{Methods}\label{methods}
\addcontentsline{toc}{section}{\protect\numberline{}Methods}

\subsection*{Descriptor Calculation}
\addcontentsline{toc}{subsection}{\protect\numberline{}Descriptor Calculation}

We consider three classes: (i) sequence--derived descriptors, (ii) aggregation descriptors, and (iii) peptide--level 3D descriptors. 
Sequence--derived descriptors --- sequence length (number of amino acids L) and net charge --- are computed directly from the amino acid sequence in FASTA format. At neutral pH, the net charge is calculated as
\begin{equation}
    q \;=\; \sum_{l=1}^{L} q_l,
    \qquad
    q_l \;=\;
    \begin{cases}
        +1, & \text{if residue } l \in \{\mathrm{Lys},\ \mathrm{Arg}\},\\[2pt]
        -1, & \text{if residue } l \in \{\mathrm{Asp},\ \mathrm{Glu}\},\\[2pt]
        0,  & \text{otherwise.}
    \end{cases}
\end{equation}

Aggregation descriptors --- aggregation propensity and SA/no-SA labels --- are taken as provided by the merged source datasets. 

All peptide--level 3D descriptors are computed from the lowest-energy structure predicted by PEP-FOLD~\cite{reyPEPFOLD4PHdependentForce2023}. The code was executed in parallel and returns the top five models per peptide; it then selects the model with the best internal score in PDB format. Failed predictions were retried up to three times. From each selected PDB, we parsed backbone and side-chain coordinates using Biopython's PDBParser~\cite{cockBiopythonFreelyAvailable2009}. Residue secondary structure is assigned with the DSSP algorithm~\cite{hekkelmanMkdsspCalculateSecondary}. Peptides for which DSSP or PDB coordinate parsing failed were omitted from conditioning and, consequently, from model training/validation --- unless AP or SA/no-SA labels were defined; in that case, the masking mechanism handles the missing proxy descriptors. All descriptor--extraction steps were parallelized.
We compute the monomeric $\beta$-strand assignment as
\begin{equation}
    f_\beta \;=\; \frac{L_E}{L},
\end{equation}
where $L_E$ is the number of residues labeled \texttt{E} (extended strand) by DSSP. Because nonzero $f_\beta$ values are rare in our dataset, we binarized this feature as a flag indicating $\beta$-strand content if $f_\beta>0$.

For the hydrophobic moment, we use the Eisenberg hydrophobicity scale $h_l$ for residue $l$~\cite{eisenbergAnalysisMembraneSurface1984} and define $\hat{\mathbf{v}}_l$ as the vector along $C_\alpha \to C_\beta$ for residue $l$ (for Gly or missing $C_\beta$, a unit vector is used). The hydrophobic-moment vector and its magnitude are
\begin{equation}
\boldsymbol{M} \;=\; \frac{1}{L}\sum_{l=1}^{L} h_l\,\hat{\mathbf{v}}_l,
    \qquad
    M \;=\; \;\sqrt{M_x^2+M_y^2+M_z^2} ,
\end{equation}
where $h_l$ weights each directional contribution $\hat{\mathbf{v}}_l$. Normalizing by $L$ yields a length-independent measure. The magnitude of the vector is then used as the final descriptor.
    
\subsection*{Aggregation Classifier and Regressor}
\addcontentsline{toc}{subsection}{\protect\numberline{}Aggregation Classifier and Regressor}

We fine-tune a lightweight, two-head predictor on top of an ESM-2 encoder (t12/35M)~\cite{linLanguageModelsProtein2022a}. Given an input peptide $y=(y_1,\dots,y_L)$, with $L_{\max}=10$, where each $y_l$ is an amino acid symbol, we tokenize the input for ESM-2 and take the final ESM layer representation, mask out special and padding tokens, and compute a masked mean sequence representation. A linear projection maps this representation to a shared hidden state, from which two heads produce (i) a scalar aggregation propensity $\hat a$ and (ii) a probability $\hat p$ for assembly. The objective loss of the model is a sum of a binary cross-entropy term on $\hat p$ and a mean-squared error term on $\hat a$. We do training in two stages: (i) firstly, we freeze the ESM-2 encoder and train only the heads; (ii) then, we unfreeze and fine-tune end-to-end with discriminative learning rates. As such, we firstly train the heads with AdamW (lr $10^{-3}$) for 5 epochs; we then unfreeze the ESM-2 encoder and fine-tune with discriminative learning rates (encoder $10^{-5}$, heads $10^{-3}$) and an exponential learning rate decay for 6 more epochs.

\subsection*{PepMorph Generative Model}
\addcontentsline{toc}{subsection}{\protect\numberline{}PepMorph Generative Model}

We model peptide sequences via a conditional variational autoencoder with a masking mechanism that supports partial specification of design descriptors. With $y=(y_1,\dots,y_L)$ denoting the peptide, where each $y_l$ is an amino acid symbol, we draw the tokenized input from the ESM-2 alphabet (20 residues plus special \texttt{<bos>}, \texttt{<eos>}, and \texttt{<pad>} tokens). We cap peptide length at $L_{\max}$ residues and form fixed-length token sequences by framing them with \texttt{<bos>} and \texttt{<eos>} and right-padding with \texttt{<pad>}. Let $c\in\mathbb{R}^d$ be the vector of $d=6$ normalized descriptors, and let $m\in\{0,1\}^d$ indicate which descriptors are available for a given sample. 

The model then leverages an encoder, a masked conditional prior, and a decoder. First, tokens are embedded and summed with sinusoidal positional encodings, then passed through a 2-layer Transformer encoder (hidden size 256, 8 heads). We apply padding masks and average the contextual embeddings across non-\texttt{<pad>} positions to obtain a fixed-dimensional sequence representation, from which two linear heads produce the mean and log-variance of a diagonal-Gaussian posterior $q(z\mid y)$ over the latent $z$. The masked conditional prior forms the condition summary $s$ as described previously. Finally, a 2-layer Transformer decoder (hidden size 256, 8 heads) autoregressively models $p(y\mid z,s)$: its cross-attention memory concatenates two tokens --- a latent token obtained by projecting $z$ and a condition token obtained by projecting $s$ --- and at each step the decoder attends to both while predicting the next amino acid symbol until emitting \texttt{<eos>}.

For a mini-batch of size $N$, we seek to minimize
\begin{equation}
    \mathcal{L}
= \mathcal{L}_{\mathrm{rec}}
+ \beta\,\mathcal{L}_{\mathrm{KL}}
+ \lambda_{\text{mask}}\mathcal{L}_{\text{mask}}
+ \lambda_{\text{bin}}\mathcal{L}_{\text{bin}}
+ \lambda_{\text{cont}}\mathcal{L}_{\text{cont}}
\end{equation}
with $\beta$ a cyclic Kullback-Leibler (KL) weight and $\lambda_{\text{mask}},\lambda_{\text{bin}},\lambda_{\text{cont}}$ scalar hyperparameters. The reconstruction loss $\mathcal{L}_{\mathrm{rec}}$ is the negative log-likelihood of the ground-truth amino-acid at each non-\texttt{<pad>} position under the decoder's categorical distribution (token-level cross-entropy with label smoothing and teacher forcing).

For the divergence term, we follow the VAEAC approach~\cite{ivanovVariationalAutoencoderArbitrary2018} and use the closed‐form KL divergence between diagonal Gaussians. Let the encoder posterior for sample $n$ be $q_n(z\mid y)=\mathcal{N}(\mu_n,\operatorname{diag}(\sigma_n^2))$ and the masked conditional prior be $p_n(z\mid s)=\mathcal{N}(\mu_{\mathrm{prior},n},\operatorname{diag}(\sigma_{\mathrm{prior},n}^2))$, both in latent dimension $K$. The KL loss is then
\begin{equation}
    \mathcal{L}_{\mathrm{KL}}
= \frac{1}{N}\sum_{n=1}^{N}
\frac{1}{2}\sum_{k=1}^{K}\!\left[
\log\frac{\sigma_{\mathrm{prior},n,k}^{2}}{\sigma_{n,k}^{2}}
+ \frac{\sigma_{n,k}^{2}+\big(\mu_{n,k}-\mu_{\mathrm{prior},n,k}\big)^{2}}{\sigma_{\mathrm{prior},n,k}^{2}}
- 1
\right].
\end{equation}

To encourage the summary $s$ to encode which descriptors are present and their values, we add three reconstruction terms. For the mask head, let $\hat m_{n,i}$ be the predicted logit for mask entry $m_{n,i}\in\{0,1\}$ (for descriptor index $i=1,\dots,d$). We define
\begin{equation}
    \mathcal{L}_{\text{mask}}
= \frac{1}{N}\sum_{n=1}^{N}\frac{1}{d}\sum_{i=1}^{d}
\Big(
-\, m_{n,i}\,\log \sigma(\hat m_{n,i})
- (1-m_{n,i})\,\log\!\big(1-\sigma(\hat m_{n,i})\big)
\Big).
\end{equation}
For the binary descriptors, let $\mathcal{B}_n=\{i\in \{1,\dots,d\}:m_{n,i}=1\}$ denote the binary descriptors that are observed for sample $n$, with target $b_{n,i}\in\{0,1\}$ and predicted logit $\hat b_{n,i}$. We define
\begin{equation}
    \mathcal{L}_{\text{bin}}
= \frac{1}{N}\sum_{n=1}^{N}\frac{1}{|\mathcal{B}_n|}\sum_{i\in\mathcal{B}_n}
\Big(
-\, b_{n,i}\,\log \sigma(\hat b_{n,i})
- (1-b_{n,i})\,\log\!\big(1-\sigma(\hat b_{n,i})\big)
\Big),
\end{equation}
i.e., the same binary cross-entropy, but applied only to the observed binary descriptors for each sample.
For the continuous descriptors, let $\mathcal{C}_n=\{i\in \{1,\dots,d\}:m_{n,i}=1\}$ be the continuous descriptors observed for sample $n$, with target value $c_{n,i}$ and prediction $\hat c_{n,i}$. We define
\begin{equation}
    \mathcal{L}_{\text{cont}}
= \frac{1}{N}\sum_{n=1}^{N}\frac{1}{|\mathcal{C}_n|}\sum_{i\in\mathcal{C}_n}
\big(\hat c_{n,i}-c_{n,i}\big)^{2}.
\end{equation}

To maximize the performance of the model and better attain the main goals of generation --- broad exploration of sequence space and strong compliance with arbitrary partial conditions ---, we introduce three adjustments to the training data usage. Firstly, we attenuate class imbalance in the binary descriptors with weighted sampling ($\times 2$ for positive \texttt{is\_assembled}, $\times 10$ for positive \texttt{has\_beta\_strand}). Adding to this, in order to expose the model to partial conditions, we apply stochastic masking during training: for each example, a random subset of currently available descriptors is set to unobserved ($m_i=0$), ensuring at least one descriptor remains observed. Finally, each epoch is augmented with $5,000$ uniformly sampled short peptides (up to $L_{\max}$) that observe only the length descriptor, seeking to achieve robustness at encoding the entirety of the peptide space.

We train for $250$ epochs using AdamW (learning rate $10^{-3}$, weight decay $10^{-4}$) and halve the learning rate on validation-loss plateaus (patience $30$ epochs). All model training was implemented in PyTorch and run on NVIDIA RTX 3090 GPUs.  The data was stratified by peptide length and split into train ($80\%$), validation ($10\%$), and test ($10\%$) sets using scikit-learn's \texttt{train\_test\_split}.

\subsection*{Conditional Likelihood Evaluation}
\addcontentsline{toc}{subsection}{\protect\numberline{}Conditional Likelihood Evaluation}

To quantify conditional generalization of the decoder, we compute teacher-forced token-level negative log-likelihood (NLL) on a held-out test set under different conditioning masks. For each test peptide $y=(y_1,\dots,y_L)$ with available descriptor vector $c$ and mask $m$, we form the condition summary $s=\phi([c\odot m,\,m])$ and run the decoder in teacher-forcing mode, i.e., at step $t$ the model is given the ground-truth prefix $(y_1,\dots,y_{t-1})$ and predicts a categorical distribution over $y_t$. We compute the per-token cross-entropy (in nats) over non-\texttt{<pad>} tokens as
\begin{equation}
\mathrm{CE}(y;c,m)
= -\frac{1}{L'} \sum_{t \in \mathcal{T}}
\log p_\theta\!\big(y_t \mid y_{<t},\, s\big),
\end{equation}
where $\mathcal{T}$ indexes all non-\texttt{<pad>} target positions (including \texttt{<eos>}) and $L'=|\mathcal{T}|$. We report the mean CE across the test set, and its corresponding perplexity $\mathrm{PPL}=\exp(\mathrm{CE})$.

We evaluate three mask settings: (i) observed-mask, using each test sample's native availability mask $m$ (reflecting naturally missing descriptors); (ii) full-mask on the complete-case subset (dataset subset where all descriptors are observed, $m=\mathbf{1}$); and (iii) random-on-top on the same complete-case subset, where we additionally drop each descriptor independently with probability $0.5$ (while ensuring at least one descriptor remains observed) and compute CE under the resulting mask.

\subsection*{Coarse-Grained Molecular Dynamics Simulations}
\addcontentsline{toc}{subsection}{\protect\numberline{}Coarse-Grained Molecular Dynamics Simulations}

Individual PDB files were built from the FASTA sequences via Avogadro v1.2~\cite{hanwellAvogadroAdvancedSemantic2012} using geometry optimization. All CG-MD simulations used GROMACS~\cite{abrahamGROMACSHighPerformance2015} with MARTINI 3~\cite{souzaMartini3General2021} forcefield (except the 16 simulations for AP control verification, that leveraged MARTINI 2.2~\cite{talluriDiscoveryUnconventionalNonintuitive2025}), applying bead substitutions (Q5 to Q4, TC5 to SC4) following Sasselli and Coluzza's adaptation to small peptides~\cite{sasselliAssessmentMARTINI32024}.  Topologies were generated with \texttt{martinize2}~\cite{kroonMartinize2VermouthUnified2025} from the all‐atom PDBs to obtain the corresponding coarse-grained representations. 

Each system comprised $300$ peptides in a $15\times15\times15~\mathrm{nm}^3$ cubic, periodic box~\cite{wangAggregationRulesShort2024}, except the 16 simulations for box-size effects, which utilized a $24\times24\times24~\mathrm{nm}^3$ box, with the same concentration ($1230$ peptides). The boxes were then solvated with the MARTINI water model and neutralized by adding Na$^{+}$/Cl$^{-}$ when needed to achieve electroneutrality. Energy minimization employed the steepest‐descent integrator for $5,000$ steps, with reaction‐field electrostatics and $1.2$ nm cutoffs for both Coulomb and Van der Waals (VdW) interactions. Production simulations consisted of 20 million steps ($\Delta t$ = 25 fs, totalizing 500 ns) in the NPT ensemble at 303 K and 1 bar. A frame checkpoint of the trajectory is stored every 50k steps (1.25 ns). We used the velocity‐rescale thermostat, C-rescale barostat and Particle--Mesh Ewald for electrostatics with a cutoff of $1.2$ nm. VdW interactions used a 1.2 nm cutoff with the potential-shift-Verlet modifier. Neighbor search used the Verlet scheme with a 0.005 buffer tolerance.
Each system was run 3 times with different initial velocities for statistical purposes. All simulation runs used GPU acceleration. 

\subsection*{All-Atom Molecular Dynamics Simulations}
\addcontentsline{toc}{subsection}{\protect\numberline{}All-Atom Molecular Dynamics Simulations}

We performed two targeted AA self-assembly follow-up simulations for IVLFIF and VFFIF using GROMACS with AMBER99SB-ILDN and TIP3P water. Each system comprised 300 peptide copies in a $15\times15\times15~\mathrm{nm}^3$ cubic periodic box. The boxes were solvated and neutralized to electroneutrality with K$^+$/Cl$^-$. Energy minimization employed steepest descent (50,000 steps) with a Verlet cutoff scheme, Particle--Mesh Ewald electrostatics (1.0~nm real-space cutoff), and a 1.0~nm van der Waals cutoff. Equilibration consisted of 100~ps NVT followed by 100~ps NPT with position restraints on the peptides, using the velocity-rescale thermostat (300~K) and the Parrinello--Rahman barostat (1~bar, isotropic). Production simulations were run in the NVT ensemble for $1.25~\mu\mathrm{s}$ with a 2~fs timestep ($n_\mathrm{steps}=6.25\times 10^{8}$), using PME electrostatics (1.0~nm) and a 1.0~nm van der Waals cutoff; coordinates were saved every 10~ps.

\subsection*{Aggregation and Morphology Metrics}
\addcontentsline{toc}{subsection}{\protect\numberline{}Aggregation and Morphology Metrics}

With the full trajectories, we can compute the SASA per saved frame with GROMACS (\texttt{gmx sasa}) on the peptide group. Let $S_t$ be the total SASA at frame $t$. We define the early-time mean $\bar S_{\mathrm{first2}}=(S_1+S_2)/2$ and the late-time mean $\bar S_{\mathrm{last2}}=(S_{T-1}+S_T)/2$, so that the simulation AP is
\begin{equation}
    \mathrm{AP} \;=\; \frac{\bar S_{\mathrm{first2}}}{\bar S_{\mathrm{last2}}}.
\end{equation}

For RMOI, we identify, for the final saved frame of the trajectory, the largest peptide aggregate using a periodic-boundary, surface-to-surface connectivity graph. Each coarse-grained bead $j$ is assigned a radius $r_j$ and a mass $m_j$ (by MARTINI bead class, using defaults~\cite{souzaMartini3General2021}). Two beads $j,k$ are considered connected if their minimum-image center distance $d_{jk}$ satisfies 
\begin{equation}
    d_{jk} - (r_j + r_k) \;<\; r_{\mathrm{cut}},
\end{equation}
with cutoff $r_{\mathrm{cut}}$ of 0.6nm. Bead coordinates are unwrapped by the minimum-image convention and translated so the mass-weighted center of mass (COM) is at the origin
\begin{equation}
    M = \sum_j m_j,\qquad
    \mathbf r_{\mathrm{COM}}=\frac{1}{M}\sum_j m_j\,\mathbf r_j,\qquad
    \tilde{\mathbf r}_j=\mathbf r_j-\mathbf r_{\mathrm{COM}}.
\end{equation}
The inertia tensor $\mathbf{I}$ about the COM is the standard mass-weighted second-moment matrix
\begin{equation}
    \mathbf{I} \;=\; \sum_j m_j\!\left(\,\|\tilde{\mathbf{r}}_j\|^2 \mathbb{I} \;-\; \tilde{\mathbf{r}}_j \tilde{\mathbf{r}}_j^{\!\top}\right).
\end{equation}
We then calculate RMOI as the ratio of the smallest and largest eigenvalues of $\mathbf{I}$ (Eq.~\ref{eq:RMOI}).

\subsection*{Statistical Analysis}
\addcontentsline{toc}{subsection}{\protect\numberline{}Statistical Analysis}

For screening-yield comparisons (hits under a finite sampling budget), we report binomial proportions with two-sided Wilson score 95\% confidence intervals. Screening efficiency is summarized by calls per hit defined as $N/H$, where $N$ is the number of generated sequences and $H$ is the number of proxy-compliant hits (reported as $\infty$ when $H=0$).

To compare PepMorph candidates against aggregation-matched descriptor-fail control cohorts, we use two-sided Fisher exact tests on $2\times 2$ contingency tables (success/failure by cohort) and report the corresponding odds ratios. When a zero cell occurs, odds ratios are interpreted qualitatively; exact $p$-values from Fisher's test are reported.


\begin{acknowledgement}

We thank the early access to the SuperMUC-NG Phase 2 cluster, allowing us to do extensive control CG-MD simulations, which considerably strengthened our work.

Funded by the European Union. Views and opinions expressed are however those of the author(s) only and do not necessarily reflect those of the European Union or the European Research Council Executive Agency. Neither the European Union nor the granting authority can be held responsible for them. This work was funded by the ERC (StG SupraModel) - 101077842.

\end{acknowledgement}

\section*{Data Availability}
The curated PepMorph dataset, as well as the resulting simulation trajectories used for \textit{in silico} validation, are made publicly available at \href{https://github.com/tummfm/pepmorph}{https://github.com/tummfm/pepmorph}.

\section*{Code Availability}
Weights for all trained and reported models, as well as code regarding the trained models and the simulation setups for \textit{in silico} validation are made publicly available at \href{https://github.com/tummfm/pepmorph}{https://github.com /tummfm/pepmorph}.

\section*{Author contributions}
\textbf{Costa, N.}: Conceptualization, Methodology, Software, Validation, Formal Analysis, Investigation, Resources, Data Curation, Writing - Original Draft, Visualization. \textbf{Zavadlav, J.}: Writing - Review \& Editing, Supervision.


\begin{suppinfo}

The following files are available free of charge.
\begin{itemize}
  \item \texttt{supplementary\_material.pdf}: Supplementary material, containing further dataset analysis and composition plots; all hyperparameters of the models used; analysis of the generation, including sampling costs and efficiency curves; per-peptide CG-MD visual snapshots and corresponding metrics and assessment of all runs; description and analysis of the robustness cohorts, including CG-MD visual snapshots and corresponding metrics and assessment of all runs; description and results of the executed atomistic simulations.
\end{itemize}

\end{suppinfo}

\bibliography{achemso-bib}
\addcontentsline{toc}{section}{\protect\numberline{}References}

\end{document}